\documentclass[notitlepage,a4,10.5pt]{article}

\usepackage{graphicx}

\usepackage{amsmath}%
\usepackage{amsfonts}%
\usepackage{amssymb}%
\usepackage{mathrsfs}
\usepackage{amsthm}
\usepackage{float}

\usepackage{authblk}

\usepackage{array}

\usepackage[left=2.5cm,right=2.5cm,top=3cm,bottom=3cm]{geometry}

\newcommand{\mc}{\mathcal}

\newcommand{\bol}{\boldsymbol}

\newcommand{\abs}[1]{\left\lvert{#1}\right\rvert}
\newcommand{\w}{\wedge}
\newcommand{\lr}[1]{\left({#1}\right)}

\newcommand{\mf}{\mathfrak}
\newcommand{\p}{\partial}

\newcommand{\ti}[1]{\textit{#1}}
\newcommand{\tb}[1]{\textbf{#1}}

\begin{document}

\title{The Effect of Spacetime Curvature on Statistical Distributions}
\author[1]{Naoki Sato} 
\affil[1]{Graduate School of Frontier Sciences, \protect\\ The University of Tokyo, Kashiwa, Chiba 277-8561, Japan \protect\\ Email: sato\_naoki@edu.k.u-tokyo.ac.jp}
\date{\today}
\setcounter{Maxaffil}{0}
\renewcommand\Affilfont{\itshape\small}

    \maketitle
    \begin{abstract}
		The Boltzmann distribution of an ideal gas is determined by the Hamiltonian function generating single particle dynamics. 
		Systems with higher complexity often exhibit topological  constraints, which are independent of the Hamiltonian and may affect the shape of the distribution function as well.  
		Here, we study a further source of heterogeneity, the curvature of spacetime arising from the general theory of relativity.
		The present construction relies on three assumptions: 
		first, the statistical ensemble is made of particles obeying geodesic equations, which define the phase space of the system. 
		Next, the metric coefficients are time-symmetric,  
		implying that, if thermodynamic equilibrium is achieved, 
		all physical observables are independent of coordinate time. 
		Finally, ergodicity is enforced with respect to proper time, so that
		 ambiguity in the choice of a time variable for the statistical ensemble is removed. 
		Under these hypothesis, we derive the distribution function of thermodynamic equilibrium, and verify that it reduces to the Boltzmann distribution in the classical limit. 		
		We further show that spacetime curvature affects 
		physical observables, even far from the source of the metric. 
	 	Two examples are analyzed: an ideal gas in Schwarzschild spacetime and a charged gas in Kerr-Newman spacetime. 
		In the Schwarzschild case, conservation of macroscopic constraints, such as angular momentum, 
		combined with relativistic distortion of the distribution function can produce configurations with decreasing density and growing azimuthal rotation velocity far from the event horizon of the central mass. 
		In the Kerr-Newman case, 
		it is found that kinetic energy associated with 
		azimuthal rotations is an increasing function of the radial coordinate, and it eventually approaches a constant value corresponding to classical equipartition, even though spatial particle density decreases. 
		\end{abstract}

\section{Introduction}
The purpose of the present study is to investigate how
the distribution of a statistical ensemble is modified 
if particles feel the spacetime curvature 
arising from the principles of general relativity. 
This problem is usually encountered in astrophysical systems, 
such as gas in proximity of a black hole, or
stars within the gravitational field of a galaxy.  

If special relativity is taken into account, 
ambiguity arises in the definition of the thermodynamic arrow of time 
with respect to which statistical processes evolve. 
This issue directly affects the notion of temperature and thermodynamic equilibrium.
Within the framework of special relativity, 
the classical Maxwell-Boltzmann distribution of an ideal gas
can be generalized through the Maxwell-J\"uttner distribution \cite{Jutt,Chacon}.
The Maxwell-J\"uttner distribution describes  
the type of statistics measured by an observed 
pinned into the coordinate frame $\lr{t,x,y,z}$ made of coordinate time $t$ and
Cartesian coordinates $\lr{x,y,z}$. 
The special relativistic particle energy, which is the constant of motion arising from the
time-symmetry of the geodesic Hamiltonian, fluctuates when particles interact. 
Here, both stochasticity of microscipic interactions and 
ergodicity 
are defined with respect to coordinate time $t$, which therefore represents  
the thermodynamic arrow of time
\cite{Cubero,Dunkel}.  
Other special relativistic generalizations of Boltzmann statistics  
using $t$ as time variable have also been proposed (see e.g. \cite{Kaniadakis}, where the collision
operator of a Boltzmann-type kinetic equation is formulated in consistency with the additivity law for relativistic momenta).

When the full framework of general relativity is considered, 
an additional difficulty emerges because the
metric coefficients themselves are subject to statistical fluctuations, 
and the equivalence between inertial mass and energy 
makes the classical understanding of thermodynamic temperature ineffective (one example of nontrivial relationship between temperature and  
structure of spacetime is the proportionality law between entropy and the area of a causal horizon \cite{Jacobson}). 
Since energy in the form of heat has inertial mass,
heat is accelerated in a gravitational field, leading to relativistic effects 
such as the Tolman-Ehrenfest law according to which in a perfect fluid at equilibrium temperature is higher at lower values of the gravitational potential energy \cite{Tolman,Frankel2}. 
It has been suggested that a relativistic temperature 
may be introduced through the notion of thermal time, which is the time-parameter associated with the
flow generated by the Hamiltonian $\log f$, where $f$ is the statistical distribution of the system \cite{Rovelli2011,Rovelli2013,Rovelli1993}. 
This definition reproduces the Tolman-Ehrenfest law $T\propto \abs{\xi}^{-1}$, with $\abs{\xi}$ the norm of a 
timelike Killing field $\xi$ associated with a stationary spacetime metric.

The classical construction of statistical mechanics hinges upon the Hamiltonian nature
of unperturbed single particle dynamics. 
Despite its ideal nature, Hamiltonian mechanics represents the 
building block of kinetic theory and statistical mechanics due to the 
volume preserving property of Hamiltonian flows described by Liouville's theorem. 
Indeed, for a canonical Hamiltonian system, the phase space volume spanned by canonical coordinates
defines an invariant measure. 
Such invariant measure originates from the symplectic structure of the phase space,
and it does not depend on the specific form of the Hamiltonian function (energy) of
a single particle. Hence, although energy will fluctuate once particles are allowed
to interact (collide) and increase the entropy of the system, the underlying 
phase space structure persists, providing the necessary condition 
to postulate the ergodic hypothesis \cite{Moore,Jaynes}. 
This formulation holds for noncanonical Hamiltonian systems \cite{LittleJ} as well,  
although noncanonicality introduces new sources of heterogeneity \cite{Yos2014}. 
Indeed, the Boltzmann distribution is achieved not in the whole phase space, 
but on each submanifold corresponding to a level set of the Casimir invariants spanning the kernel of the 
noncanonical Poisson bracket. On a Casimir leaf, 
canonical coordinates and the associated Liouville measure can be locally recovered by application of the Lie-Darboux theorem \cite{Arnold,deLeon}.  
Ergodicity is therefore invoked on each Casimir leaf, 
and equilibrium is expressed by a generalized Boltzmann distribution, 
which is an explicit function of the single particle Hamiltonian and the Casimir invariants,
and it is related in a nontrivial manner to the density of states in the 
dynamical variables of the original noncanonical form \cite{SatoPRE,SatoPhysD20}. 
In a similar way, the approach toward relativistic statistical mechanics developed in this paper 
will be based on the Hamiltonian structure of the geodesic equations of motion (for a discussion of the 
proper setting of relativistic kinetic theory see e.g. \cite{Mis}).

There are at least three different levels of relativistic Hamiltonian dynamics that one could exploit
to formulate a statistical theory. The first type of Hamiltonian structure is that associated with 
special relativity. Here, the single particle Hamiltonian $\mf{H}=m\gamma c^2$ is the generator of dynamics 
($m$ is the particle rest mass and $\gamma=dt/d\tau$ the Lorentz factor) and the equations of motion
express the evolution of Cartesian coordinates $\lr{x,y,z}$ with respect to coordinate time $t$. 
The Hamiltonian $\mf{H}$ can be identified with (minus) the constant of motion $p_0$ arising from the time-symmetry $\p_0H=0$ of the
geodesic Hamiltonian $H=g^{ij}p_ip_j/2m$, with $g^{ij}$ the contravariant metric tensor and $p_i$, $i=1,...,4$,
the canonical momenta associated with spacetime coordinates $\lr{x^0,x^1,x^2,x^3}=\lr{t,x,y,z}$.  
In this setting, stochastic interactions among particles change the value of $\mf{H}$ 
by breaking the time-symmetry. This process leads to the Maxwell-J\"uttner distribution, which is
an equilibrium state $f_t\lr{x^1,x^2,x^3,p_1,p_2,p_3}$ depending only on 
spatial coordinates and momenta and defined with respect to coordinate time $t$. 
This approach can be generalized beyond the Minkowski metric, as long as
the time-symmetry of the geodesic Hamiltonian guarantees the existence of the constant of motion $p_0$ (see for example \cite{Kim}, where the case of the Rindler metric is analyzed).  

In the second setting, which is the one studied in the present paper, phase space is assigned by the geodesic equations of motion with Hamiltonian $H$, although the time-symmetry of $H$ is assumed to hold. 
Here, it is the geodesic Hamiltonian $H$ (and not the constant of motion $p_0=-\mf{H}$) the 
physical quantity subject to statistical fluctuations resulting from particle collisions, 
and the thermodynamic arrow of time is given by proper time $\tau$. 
Notice that particle interactions are assumed to preserve the time-symmetry so that 
the constant of motion $p_0$ survives collisions and effectively behaves as the Casimir invariant 
of a noncanonical Hamiltonian system with dynamical variables $\lr{x^1,x^2,x^2,p_0,p_1,p_2,p_3}$
evolving in proper time $\tau$. 
The expected result at thermodynamic equilibrium is a Boltzmann-type distribution $f_{\tau}\lr{x^1,x^2,x^2,p_1,p_2,p_3;p_0}$ on the phase space submanifold defined by a level set of $p_0$, which behaves as an external parameter labeling the foliation.

In the third scenario, time-symmetry does not hold anymore, and the Hamiltonian structure 
is provided by the geodesic Hamiltonian $H$, which generates a proper time flow  
in $8$-dimensional phase space with canonical coordinates $\lr{x^0,x^1,x^2,x^3,p_0,p_1,p_2,p_3}$.
The outcome is an equilibrium state $f_{\tau}\lr{x^0,x^1,x^2,x^3,p_0,p_1,p_2,p_3}$.
Notice that, since the arrow of time is given by proper time and time-symmetry is absent, 
thermodynamic equilibrium is expressed as a function of coordinate time $x^0$ and momentum $p_0$, 
which are treated on the same ground of the other coordinates. This third case is not discussed in the
present study. 
Table \ref{tab1} summarizes the three scenarios above. 

\begin{table}[h]
\centering
\begin{tabular}{|c|c|c|c|}
\hline
\multicolumn{4}{|c|}{Relativistic statistical mechanics} \\
\hline
Hamiltonian & Time & Symmetry& Equilibrium\\
\hline
$-p_0$&$t$&$\p_0$&$f_{t}\lr{x^1,x^2,x^3,p_1,p_2,p_3}$\\
\hline
$H=g^{ij}p_ip_j/2m$&$\tau$&$\p_0$&$f_{\tau}\lr{x^1,x^2,x^3,p_1,p_2,p_3;p_0}$\\
\hline
$H=g^{ij}p_ip_j/2m$&$\tau$&$\times$&$f_{\tau}\lr{x^0,x^1,x^2,x^3,p_0,p_1,p_2,p_3}$\\
\hline
\end{tabular}
\caption{\label{tab1} Hamiltonian structure of single particle dynamics, thermodynamic arrow of time, spacetime symmetry, and equilibrium distribution function for three relativistic dynamical settings.}
\end{table}

We remark that non-gravitational forces can be included in the theory by modifying the geodesic Hamiltonian. 
Furthermore, in principle each particle may contribute to the spacetime metric by appropriate coupling with the 
Einstein field equations \cite{Wald}, although this is not pursued here. 
The evolution equation for the distribution function of the system can be derived once 
the Hamiltonian structure associated with single particle dynamics and the symmetries of particles collisions have been assigned. 
Such collisions, which are intended as the set of all sources inducing fluctuations in the single particle Hamiltonian function, 
are assumed to occur on time scales enabling the thermalization of the system (see e.g. \cite{Lynden,Chavanis} on the problem of  collisionless relaxation in stellar systems).
Here, both the Hamiltonian structure and the properties of particle collisions are chosen on physical grounds. 
Notice that, once given, the Hamiltonian structure also determines the thermodynamic arrow of time. 
The procedure to derive a Fokker-Planck-type evolution equation for the distribution function 
associated with a given Hamiltonian structure can be found in reference \cite{SatoPhysD20}, 
while in \cite{Smerlak} a Fokker-Planck equation consistent with the Tolman-Ehrenfest law is constructed 
by taking into account the curvature of spacetime. 
We also refer the reader to \cite{SatoJMatPhys} for an example 
outside the context of relativity (a system of nonholonomically constrained charged particles) 
where proper time  is used as time variable to characterize the evolution of a statistical ensemble.

Our aim in this paper is to understand how a nontrivial spacetime metric
affects the spatial distribution of matter and its macroscopic properties at thermodynamic equilibrium, 
and to clarify whether general relativistic effects persist at large distances from their source. 
If true, this latter fact implies that a classical statistical description is not physically sound even if 
local single particle dynamics is effectively classical due to separation from the source of spacetime distortion. 
This problem has practical implications: for example, the effect of spacetime curvature on the values of macroscopic observables such
as the average rotation speed of matter around the galactic center represents a property of interest in astrophysical studies 
concerning unconventional matter (see e.g. \cite{Balasin,Crosta,Cooperstock}).  
As already mentioned, we consider the setting of case 2 in table \ref{tab1}. 
This configuration is appropriate to describe a system 
where the constant of motion $p_0$ is not broken by particle collisions.
For a special relativistic particle $p_0=-m\gamma c^2=-mc^2 dt/d\tau$, implying that
during a collision the individual clock speed $dt/d\tau$ is unaltered, although the energy $H$ is.  
The time-symmetry $\p_0$ then makes it possible to achieve equilibrium states that are
independent of coordinate time $t$. Such states express the probability density of finding
a particle in a given region of the phase space in the proper time limit $\tau\rightarrow\infty$.

The present paper is organized as follows. 
In section 2, first the geodesic equations of motion 
are reduced to a 6-dimensional canonical Hamiltonian system 
on a level set of the constant of motion $p_0$ arising from the time-symmetry of the metric coefficients.
Then, the analogous of the Boltzmann distribution is derived 
by using the Liouville measure in the reduced phase space, and 
the spatial density distribution at thermodynamic equilibrium is obtained
by integrating the distribution function in momentum space. 
Such density is found to be distorted by the determinant of the spatial part of the metric tensor 
(which is related to the Riemannian curvature of the corresponding 3-manifold), and an exponential factor
involving the spacetime coefficients of the metric tensor. Hence, the higher the inhomogeneity of 
spacetime curvature, the higher the inhomogeneity of the spatial distribution at thermodynamic equilibrium.  
In section 3, we discuss the relationship among the distribution function derived in section 2,
the special relativistic Maxwell-J\"uttner distribution, and the classical Maxwell-Boltzmann distribution.
In section 4, the theory is applied to the case of an ideal gas lying in Schwarzschild's exterior spacetime \cite{Schw}. 
In particular, spatial density distribution and average azimuthal rotation velocity are evaluated explicitly. 
It is shown that the combination of macroscopic constraints, such as conservation of angular momentum, 
with relativistic distortion of the statistical distribution may result in a decreasing spatial density distribution
and a growing azimuthal rotation velocity faraway from the event horizon of the central mass generating the spacetime metric.  
In section 5, a similar analysis is carried out for the rotational kinetic energy associated with Kerr-Newman spacetime \cite{Kerr,Carter,Tauk,Newman1965,Boyer}. 
In this case, density decreases with radius, and, even in the absence of macroscopic constraints, 
the kinetic energy of azimuthal rotations becomes an increasing function of the radial coordinate,
eventually approaching a constant value corresponding to classical equipartition. 
Concluding remarks are given in section 6. 
 
Finally, notice that the examples studied in sections 4 and 5 rely on a number of physical parameters expressing 
the properties of the source responsible for the metric, and a set of Lagrange multipliers, such as
the equivalent of the classical inverse temperature $\beta$, whose value is not known a priori. 
Hence, physical observables are evaluated by exploring the parameter space in the neighborhood of unity.








\section{Statistical Equilibrium in Curved Spacetime}
We consider a universe $U$ of dimension $d=1+n$, with $n\geq 1$ a natural number. 
Usually, $n=3$. We assign coordinates $\lr{x^0,x^1,...,x^n}$, where $x^0=t$ is the time variable. 
The spacetime metric is
\begin{equation}
ds^2=\sum_{i,j=1}^dg_{ij}dx^idx^j=g_{00}dt^2+2\sum_{i=1}^ng_{0i}dtdx^i+\sum_{i,j=1}^n\mf{g}_{ij}dx^idx^j.\label{ds20}
\end{equation}
Here, $g_{ij}$ denotes the $d$-dimensional covariant metric tensor, 
and $\mf{g}_{ij}$ its $n$-dimensional sub-matrix corresponding to the coordinates $\lr{x^1,...,x^n}$.    
From this point on, summation on repeated indexes will be used, and ranges of summation will be omitted. 

In the context of general relativity, the tensor $g_{ij}$ is obtained
as solution of the Einstein's field equations.   
Let $N>>1$ be the number of identical particles populating $U$. 
Here, the word particle is used in the sense of element of a
statistical ensemble, such as a massive charged particle, 
a star, or a planet. 
Notice that, in principle, all particles in the ensemble contribute
in shaping spacetime metric and curvature.
In the absence of non-gravitational forces, the trajectory (wordline) of a particle is a geodesic associated with the metric \eqref{ds20}:
\begin{equation}
\ddot{x}^i=-\Gamma^i_{jk}\dot{x}^j\dot{x}^k,~~~~i=0,...,n.\label{geo}
\end{equation} 
Here, the dot stands for differentiation with respect to proper time $\tau=s/c$, with $c$ the speed of light, while 
\begin{equation}
\Gamma^{i}_{jk}=\frac{1}{2}g^{im}\lr{\frac{\p g_{mj}}{\p x^k}+\frac{\p g_{mk}}{\p x^j}+\frac{\p g_{jk}}{\p x^m}},~~~~i,j,k=0,...,n,\label{chri}
\end{equation}
are Christoffel symbols. 
The geodesic equation \eqref{geo} admits a canonical Hamiltonian 
representation. 
Let $m$ denote the rest mass of a particle in the ensemble.
The Hamiltonian of the system is
\begin{equation}
H=\frac{1}{2m}g^{ij}p_ip_j,\label{H}
\end{equation}
where the $p_i$, $i=0,...,n$, are the canonical momenta 
associated with the spacetime coordinates $\lr{x^0,...,x^n}$.
Then, system \eqref{geo} can be written in canonical Hamiltonian form
\begin{equation}
\dot{p}_i=-\frac{\p H}{\p x^i},~~~~\dot{x}^i=\frac{\p H}{\p p_i},~~~~i=0,...,n.
\label{geoH}
\end{equation}

For the ensemble to possess an equilibrium state independent of coordinate time $x^0$, we demand 
the metric coefficients $g^{ij}$ (and thus the Hamiltonian $H$) to be symmetric with respect to the time variable $x^0$, i.e.
\begin{equation}
\frac{\p g^{ij}}{\p x^0}=0,~~~~i,j=0,...,n.\label{sym0}
\end{equation}
Equation \eqref{sym0} combined with \eqref{geoH} implies that
the canonical momentum $p_0$ is a constant of motion. 
We shall see that $p_0$ corresponds to minus the special relativistic energy of the particle. 
On each level set of $p_0$, 
we may consider the $2n$-dimensional reduced Hamiltonian system 
with Hamiltonian
\begin{equation}
\mc{H}=\frac{1}{2m}\lr{g^{00}p_0^2+2g^{0i}p_0p_i+\mf{g}^{ij}p_ip_j},\label{H2}
\end{equation}
and canonical variables $\lr{p_1,...,p_n,x^1,...,x^n}$.
Notice that, in eq. \eqref{H2}, $p_0$ plays the role of a parameter, 
and the range of summation is $1,...,n$.    
In virtue of Liouville's theorem, 
the reduced system is endowed with the invariant measure
\begin{equation}
d\Pi=dp_1\w ...\w dp_n\w dx^1\w ... \w dx^n,\label{IM}
\end{equation}
on each each level of $p_0$. 
The invariant measure \eqref{IM} can be used to postulate 
an ergodic hypothesis. Since the volume $d\Pi$ is invariant, 
the probability density $f=f\lr{p_1,...,p_n,x^1,...,x^n}$ can be identified with 
the probability $dP=fd\Pi$ of finding a particle within $d\Pi$ at the point $\lr{p_1,...,p_n,x^1,...,x^n}$ in phase space. 
Hence, the information measure
\begin{equation}
S=-\int_{\Pi}f\log f\, d\Pi,\label{S}
\end{equation}
can be regarded as the thermodynamic entropy of the system. Here, $\Pi$
is the $2n$-dimensional domain (phase space) spanned by $\lr{p_1,...,p_n,q^1,...,q^n}$. 
In the absence of additional constraints, 
conservation of probability $N$ and total energy $E$, 
\begin{equation}
N=\int_{\Pi}fd\Pi,~~~~E=\int_{\Pi}f\mc{H}\,d\Pi,\label{NE}
\end{equation}
leads to a maximum entropy state described by the variational principle
\begin{equation}
\delta\lr{S-\alpha N-\beta E}=0,
\end{equation}
where variations are carried out with respect to $f$ and $\alpha$, $\beta$ are Lagrange multipliers. 
The result is the equilibrium distribution function of equal probability density on energy contours, 
the Boltzmann distribution
\begin{equation}
f=\frac{1}{Z}e^{-\beta\mc{H}}=\frac{1}{Z}\exp\left\{-\frac{\beta}{2m}\left(g^{00}p_0^2+2g^{0i}p_ip_0+\mf{g}^{ij}p_ip_j\right)\right\}.\label{feq}
\end{equation}
Here, $Z=e^{1+\alpha}$ is a normalization factor such that $\int_{\Pi}f\,d\Pi=1$.
The diffusion process maximizing 
the entropy $S$ and leading to the equilibrium state \eqref{feq}
can be formally constructed according to the procedure described in \cite{SatoPhysD20}. 
We shall discuss how to physically choose the value of the parameter $p_0$ later. 

Next, observe that, since the canonical momentum $p_0$, which has dimensions of energy, is a constant of motion, the reduced dynamics with Hamiltonian $\mc{H}$ 
is not affected by the addition of an arbitrary function $h=h\lr{p_0}$ to the Hamiltonian,
\begin{equation}
\mc{H}'=\mc{H}+h\lr{p_0}.\label{Hprime}
\end{equation}
The above redefinition of energy can be interpreted in the context of
noncanonical Hamiltonian mechanics: 
discarding the dynamical variable $x^0$, which does not affect the
evolution of the others, one can define a $2n+1$ dimensional noncanonical
Hamiltonian system with dynamical variables $\bol{z}=\lr{p_1,...,p_n,x^1,...,x^n,p_0}$
where $p_0$ plays the role of a Casimir invariant, i.e. a
function whose gradient belongs to the null-space of the Poisson matrix
\begin{equation}
\mc{J}=\begin{bmatrix}\bol{0}&-I&0\\I&\bol{0}&0\\0&0&0\end{bmatrix}.
\end{equation}
In this notation, $I$ and $\bol{0}$ are the $n$-dimensional identity matrix and null matrix respectively. Then, $\dot{p}_0=\nabla_{\bol{z}}p_0\cdot\mc{J}\nabla_{\bol{z}}\mc{H}=0$ $\forall$ $\mc{H}$, implying that $p_0$ is a Casimir invariant. Here, $\nabla_{\bol{z}}$ denotes the gradient with respect to the
variables $\bol{z}$. 
The redefinition of energy \eqref{Hprime} does not affect
the distribution function \eqref{feq} as well, since
any multiplying factor depending solely on $p_0$ arising from variation of the total energy $E'=E+\int_{\Pi}fh\lr{p_0}\,d\Pi$ will be absorbed by the normalization constant $Z$.

At equilibrium, the spatial particle density $\rho=\rho\lr{x^1,...,x^n}$ 
can be calculated by integrating the particle distribution function \eqref{feq} 
with respect to the momenta $p_1,...,p_n$. 
The range of integration for the momenta is the whole $\mathbb{R}^n$ (recall that relativistic momentum $p_k=m\gamma g_{ki}dx^i/dt$
is proportional to the relativistic mass $m\gamma$, and therefore diverges when velocity $\abs{d\bol{x}/dt}$ approaches the speed of light).
Defining $d^n p=dp_1\w ...\w dp_n$, we have,
\begin{equation}
\rho=\frac{1}{Z}\int_{\mathbb{R}^n}e^{-\beta\mc{H}}\,d^np=\frac{1}{Z}\exp\left\{-\frac{\beta}{2m}g^{00}p_0^2\right\}\int_{\mathbb{R}^n}
\exp\left\{-\frac{\beta}{2m}\lr{2g^{0i}p_ip_0+\mf{g}^{ij}p_ip_j}\right\}\,d^np.\label{rho1}
\end{equation}
The above integral can be evaluated explicitly as a series of $n$ Gaussian integrals.
If the coefficients $g^{0i}$, $i=1,2,3$, vanish, the result of \eqref{rho1} is simply,
\begin{equation}
\rho=\frac{1}{Z}\lr{\frac{2\pi m}{\beta}}^{n/2}\exp\left\{-\frac{\beta}{2m}g^{00}p_0^2\right\}\frac{1}{\sqrt{\abs{\mf{g}^{ij}}}},\label{rho2}
\end{equation}
where $\abs{\mf{g}^{ij}}$ is the determinant of $\mf{g}^{ij}$. 
When the cross terms $g^{0i}$, $i=1,2,3$, do not vanish, the Gaussian integrals are shifted.
Let us evaluate \eqref{rho1} explicitly for the case $n=3$ of general relativity.
Setting $Z'=Z\exp\left\{\frac{\beta}{2m}g^{00}p_0^2\right\}$, we have
\begin{equation}
\begin{split}
\rho=&\frac{1}{Z}\int_{\mathbb{R}^3}e^{-\beta \mc{H}}d^3p\\
=&\frac{1}{Z'}\int_{\mathbb{R}^2}\exp\left\{-\frac{\beta}{2m}\lr{2g^{0i}p_ip_0+\mf{g}^{ij}p_ip_j} \right\}d^3p\\
=&\frac{1}{Z'}\int_{\mathbb{R}^3}\exp\left\{-\frac{\beta}{2m}\left[
g^{11}p_1^2+2p_1\lr{g^{01}p_0+g^{12}p_2+g^{13}p_3}+g^{22}p^2_2+2p_2\lr{g^{02}p_0+g^{23}p_3}+g^{33}p_3^2+2g^{03}p_3p_0
\right]\right\}d^3p\\
=&\frac{1}{Z'}\int_{\mathbb{R}^3}\exp\left\{-\frac{\beta}{2m}\left[\lr{\sqrt{g^{11}}p_1+\frac{g^{01}p_0+g^{12}p_2+g^{13}p_3}{\sqrt{g^{11}}}}^2-\frac{\lr{g^{01}p_0+g^{12}p_2+g^{13}p_3}^2}{g^{11}}\right]\right\}\\
&\exp\left\{-\frac{\beta}{2m}\left[g^{22}p^2_2+2p_2\lr{g^{02}p_0+g^{23}p_3}+g^{33}p^2_3+2g^{03}p_3p_0\right]\right\}d^3p\\
=&\frac{1}{Z'}\sqrt{\frac{2\pi m}{\beta}}\exp\left\{\frac{\beta}{2m}\left[\frac{\lr{g^{01}}^2}{g^{11}}+
\frac{\lr{g^{02}g^{11}-g^{01}g^{12}}^2}{g^{11}\lr{g^{11}g^{22}-\lr{g^{12}}^2}}
\right]p_0^2\right\}\frac{1}{\sqrt{g^{11}}}\\
&\int_{\mathbb{R}^2}\exp\left\{-\frac{\beta}{2m}\left[
\lr{\sqrt{g^{22}-\frac{\lr{g^{12}}^2}{g^{11}}}p_2+\frac{p_3\lr{g^{11}g^{23}-g^{12}g^{13}}+p_0\lr{g^{02}g^{11}-g^{01}g^{12}}}{\sqrt{g^{11}}\sqrt{g^{11}g^{22}-\lr{g^{12}}^2}}}^2\right]\right\}
\\&\exp\left\{
-\frac{\beta}{2m}\left[\lr{g^{33}-\frac{\lr{g^{13}}^2}{g^{11}}-\frac{\lr{g^{11}g^{23}-g^{12}g^{13}}^2}{g^{11}\lr{g^{11}g^{22}-\lr{g^{12}}^2}}}p_3^2\right]
\right\}\\
&\exp\left\{-\frac{\beta}{2m}\left[2p_3p_0\lr{g^{03}-\frac{g^{01}g^{13}}{g^{11}}-\frac{\lr{g^{02}g^{11}-g^{01}g^{12}}\lr{g^{11}g^{23}-g^{12}g^{13}}}{g^{11}\lr{g^{11}g^{22}-\lr{g^{12}}^2}}}\right]\right\}dp_2dp_3\\
=&\frac{1}{Z'}\frac{2\pi m}{\beta}\exp\left\{\frac{\beta}{2m}\left[\frac{\lr{g^{01}}^2}{g^{11}}+
\frac{\lr{g^{02}g^{11}-g^{01}g^{12}}^2}{g^{11}\lr{g^{11}g^{22}-\lr{g^{12}}^2}}
+\frac{\psi^2}{\phi}\right]p_0^2\right\}\frac{1}{\sqrt{g^{11}g^{22}-\lr{g^{12}}^2}}\\
&\int_{\mathbb{R}}
\exp\left\{-\frac{\beta}{2m}\left[\lr{\sqrt{\phi}p_3+\frac{\psi}{\sqrt{\phi}}p_0}^2\right]\right\}
dp_3\\
=&\frac{1}{Z'}\lr{\frac{2\pi m}{\beta}}^{3/2}\exp\left\{\frac{\beta}{2m}\left[\frac{\lr{g^{01}}^2}{g^{11}}+
\frac{\lr{g^{02}g^{11}-g^{01}g^{12}}^2}{g^{11}\lr{g^{11}g^{22}-\lr{g^{12}}^2}}
+\frac{\psi^2}{\phi}\right]p_0^2\right\}\frac{1}{\sqrt{\abs{\mf{g}^{ij}}}}.\label{rho3}
\end{split}
\end{equation}
where, in the penultimate passage, we introduced the quantities
\begin{subequations}
\begin{align}
\phi&=g^{33}-\frac{\lr{g^{13}}^2}{g^{11}}-\frac{\lr{g^{11}g^{23}-g^{12}g^{13}}^2}{g^{11}\lr{g^{11}g^{22}-\lr{g^{12}}^2}}=\frac{\abs{\mf{g}^{ij}}}{g^{11}g^{22}-\lr{g^{12}}^2},\\
\psi&=g^{03}-\frac{g^{01}g^{13}}{g^{11}}-\frac{\lr{g^{02}g^{11}-g^{01}g^{12}}\lr{g^{11}g^{23}-g^{12}g^{13}}}{g^{11}\lr{g^{11}g^{22}-\lr{g^{12}}^2}}.
\end{align}
\end{subequations} 
From \eqref{rho3}, we thus have
\begin{equation}
\begin{split}
\rho&=\frac{1}{Z}\lr{\frac{2\pi m}{\beta}}^{3/2}\exp\left\{-\frac{\beta}{2m}\left[g^{00}-\frac{\lr{g^{01}}^2}{g^{11}}-
\frac{\lr{g^{02}g^{11}-g^{01}g^{12}}^2}{g^{11}\lr{g^{11}g^{22}-\lr{g^{12}}^2}}
\right]p_0^2\right\}\\
&\exp\left\{\frac{\beta}{2m}
\frac{\left[\lr{g^{03}g^{11}-g^{01}g^{13}}\lr{g^{11}g^{22}-\lr{g^{12}}^2}-\lr{g^{02}g^{11}-g^{01}g^{12}}\lr{g^{11}g^{23}-g^{12}g^{13}}\right]^2}{\lr{g^{11}}^2\lr{g^{11}g^{22}-\lr{g^{12}}^2}\abs{\mf{g}^{ij}}}p_0^2
\right\}\frac{1}{\sqrt{\abs{\mf{g}^{ij}}}}\\
&=\frac{1}{Z}\lr{\frac{2\pi m}{\beta}}^{3/2}\exp\left\{-\frac{\beta}{2m}\frac{\abs{g^{ij}}}{\abs{\mf{g}^{ij}}}p_0^2\right\}\frac{1}{\sqrt{\abs{\mf{g}^{ij}}}}.\label{rho4}
\end{split}
\end{equation}
Here, $\abs{g^{ij}}$ denotes the determinant of the contravariant metric tensor $g^{ij}$.
Notice that equation \eqref{rho4} reduces to \eqref{rho2} when $g^{0i}=0$, $i=1,2,3$. 
Furthermore, if the spatial part $\mf{g}^{ij}$ is diagonal, one obtains
\begin{equation}
\rho=\frac{1}{Z}\lr{\frac{2\pi m}{\beta}}^{3/2}\exp\left\{
-\frac{\beta}{2m}
\left[g^{00}-\frac{\lr{g^{01}}^2}{g^{11}}-\frac{\lr{g^{02}}^2}{g^{22}}-\frac{\lr{g^{03}}^2}{g^{33}}\right]p_0^2\right\}\frac{1}{\sqrt{\abs{\mf{g}^{ij}}}}.\label{rho5}
\end{equation}
Equation \eqref{rho4} shows that spacetime curvature affects the particle density distribution 
at thermodynamic equilibrium through the purely spatial term $1/\sqrt{\abs{\mf{g}^{ij}}}$ 
and the spacetime part contained in the exponential. 
Observe that the spatial coordinates $\lr{x^1,x^2,x^3}$ do not
need to be the usual laboratry (Cartesian) coordinates $\lr{x,y,z}$. 
Denoting with $\rho_{\rm lab}$ the particle density in the laboratory frame
and with $J$ the Jacobian determinant of the coordinate change $dx^1\w dx^2\w dx^3=Jdx\w dy\w dz$, it follows that
the density observed in the laboratory is
\begin{equation}
\rho_{\rm lab}=\rho J.
\end{equation}

At this point there are two aspects that deserve clarification 
for the distribution function $f$ of equation \eqref{feq} to make physically sense: the value of the normalization factor $Z$ and the choice of the
parameter $p_0$. 
For the purpose of the present study, we assume that
the phase space is $\Pi=\Omega\otimes \mathbb{R}^n$, where
$\Omega$ is a smooth bounded domain in $\mathbb{R}^n$. 
Then, the normalization factor $Z$ can be evaluated by recalling that $\int_{\Omega}\rho\,d^nx=1$. 
For $n=3$, equation \eqref{rho4} gives: 
\begin{equation}
Z=\lr{\frac{2\pi m}{\beta}}^{3/2}\int_{\Omega}\exp\left\{-\frac{\beta}{2m}\frac{\abs{g^{ij}}}{\abs{\mf{g}^{ij}}}p_0^2\right\} \frac{1}{\sqrt{\abs{\mf{g}^{ij}}}}d^3x. 
\end{equation}
For a sufficiently regular integrand, the above integral is well-defined.
Regarding the value of $p_0$, we argue that 
the choice $p_0=-mc^2$ is justified on physical grounds. 
To see this, let us consider the simplified case in which 
$g_{ij}=\eta_{ij}$, with $\eta_{ij}$ the metric tensor
of Minkowski spacetime (one could imagine a scenario in which all particles are initially placed
in Minkowski's flat spacetime).
Then, from Hamilton's canonical equations,
\begin{equation}
\dot{t}=\frac{\p H}{\p p_0}=-\frac{p_0}{m c^2}.
\end{equation}
Hence, $p_0$ (which is a constant of motion due to the time-symmetry  
of the Hamiltonian $H$) measures the speed of the clock associated with
the motion of a single particle through a proportionality coefficient expressing the rest energy of a particle.
If all particles are initially at rest with respect to each other,  
indistinguishability of particles
implies that all clocks are identical. 
The freedom in the choice of time units then allows one to postulate $\dot{t}=1$ at $t=0$ and thus $p_0=-mc^2$, so that the whole ensemble is effectively constrained to a $2n$-dimensional subset of  the full ($2d$-dimensional) phase space.
Nevertheless, in the following we shall not specify the chosen value for $p_0$ 
to keep the generality of the construction.
For a general metric, the relationship between $\dot{t}$ and $p_0$ is given by
\begin{equation}
\dot{t}=\frac{\p H}{\p p_0}=\frac{g^{00}p_0+g^{0i}p_i}{m}.
\end{equation}
As it will be discussed in the following sections, the constant $p_0$ arsing from the time-symmetry of the geodesic Hamiltonian can be identified
with (minus) an extension of the classical notion of particle energy. 
Then, the restriction to the value $p_0=-mc^2$ implies that all particles initially possess 
exactly the same amount of this type of energy (equivalent to their rest energy $mc^2$) 
regardless of their initial position in space. 
If the number of particles having a certain initial value of $p_0$ is given by a distribution $\sigma\lr{p_0}$,  
the present theory must be reformulated through the $p_0$-averaged distribution function
\begin{equation}
\tilde{f}=\int_{-\infty}^{-mc^2}\sigma f\,dp_0.
\end{equation} 
In this paper, we shall not pursue this possibility, but instead focus on a single level set of $p_0$. 

It is useful to briefly discuss how the theory changes if the metric coefficients are not time-symmetric, and are therefore
allowed to be explicit functions of coordinate time $x^0$.
For simplicity, we assume $n=3$ ($d=4$). 
In this case, after deriving the distribution function $f=f\lr{p_0,p_1,p_2,p_3,x^0,x^1,x^2,x^3}$ in $8$-dimensional phase space by maximization of entropy,  
the spacetime density distribution of particles $\rho=\rho\lr{x^0,x^1,x^2,x^3}$
can be obtained by integrating equation \eqref{rho4} with respect to $p_0$. The result is 
\begin{equation}
\rho=\frac{1}{Z}\lr{\frac{2\pi m}{\beta}}^2\frac{1}{\sqrt{\abs{g^{ij}}}}.
\end{equation}
The corresponding spatial density distribution seen by a stationary observer
in the proper time interval $d\tau=dt/\gamma$ is
\begin{equation}
\rho dt=\rho\gamma d\tau=\frac{c}{Z}\lr{\frac{2\pi m}{\beta}}^2\sqrt{\frac{\abs{g_{ij}}}{-g_{00}}}d\tau.\label{rhodt}
\end{equation}
Here, 
we used the facts that $\abs{g_{ij}}=1/\abs{g^{ij}}$ and $\gamma=c/\sqrt{-g_{00}}$. 
This result is reminiscent of Tolman's law for a spherical distribution of perfect fluid at equilibrium in the weak fields approximation. 
Indeed, setting $\lr{x^1,x^2,x^3}=\lr{R,\theta,\phi}$ to be spherical coordinates, due to spherical symmetry $\sqrt{\abs{g_{ij}}}=\sqrt{-g_{00}g_{11}}R^2\sin\theta$. 
If the system eventually settles to an equilibrium state independent of $x^0$, 
the value of the spatial density $u=u\lr{x^1,x^2,x^3}$ is proportional to the quantity \eqref{rhodt}.  
Hence, the proper spatial density $u_p=u/\abs{\sqrt{\mf{g}_{ij}}}$ associated with the proper volume $\sqrt{\abs{\mf{g}_{ij}}}d^3x=\sqrt{g_{11}}R^2\sin\theta d^3x$ is a spatial constant.   
Furthermore, the proper spatial mass energy density $u^{\ast}_p$, which is the sum of rest mass energy density and gravitational potential energy density, is related to $u_p$ by $u^{\ast}_{p}\sim u_{p}\sqrt{-g_{00}}$ (this is due to the relationship between $\sqrt{-g_{00}}$ and the Newtonian gravitational potential, on this point see \cite{Frankel2}). 
If we define $u^{\ast}_{p}=k/T$, with $k$ a real constant and $T$ the temperature of the system, it follows that
\begin{equation}
T\propto\frac{1}{\sqrt{-g_{00}}}.\label{TTol}
\end{equation}


\section{Relation with the Maxwell-J\"uttner and the Maxwell-Boltzmann distribution}
In this section we explore the relationship between the equilibrium distribution function \eqref{feq} 
and the Maxwell-J\"uttner distribution of special relativity.  
The classical limit leading to the usual Maxwell-Boltzmann distribution for an ideal gas is also discussed. 

In Minkowski spacetime $g_{ij}=\eta_{ij}$, the Hamiltonian \eqref{H} takes the form
\begin{equation}
H=\frac{1}{2m}\lr{-\frac{p_0^2}{c^2}+\bol{p}^2}.\label{HSR}
\end{equation} 
Here, we used the usual vector notation, $\bol{p}=\lr{p_1,p_2,p_3}$ and $\bol{x}=\lr{x^1,x^2,x^3}$. 
Thus, Hamilton's canonical equations reduce to
\begin{subequations}\label{SR}
\begin{align}
\dot{\bol{p}}&=\bol{0},\\
\dot{p}_0&=0,\\
\dot{\bol{x}}&=\frac{\bol{p}}{m},\\
\dot{t}&=-\frac{p_0}{mc^2}.
\end{align}
\end{subequations}
On the other hand, recall that the geodesic Hamiltonian \eqref{H} is proportional to  
the squared norm of the four-momentum, which satisfies $g^{ij}p_ip_j=m^2g_{ij}\dot{x}^i\dot{x}^j=-m^2c^2$. Therefore, 
\begin{equation}
H=-\frac{1}{2}mc^2.\label{HR}
\end{equation}
Here, the minus sign comes from the adopted convention on the signature of the metric tensor. 
Combining \eqref{HSR}, \eqref{SR}, and \eqref{HR}, one obtains the Lorentz factor
\begin{equation}
\dot{t}=\gamma\lr{\bol{p}}=\sqrt{1+\frac{\bol{p}^2}{m^2c^2}}.\label{LF}
\end{equation}
Using \eqref{LF}, system \eqref{SR} leads to the following equations in time $t$,
\begin{subequations}\label{SR2}
\begin{align}
\frac{d\bol{p}}{dt}=&\bol{0},\\
\frac{d\bol{x}}{dt}=&\frac{\bol{p}}{m\gamma}.
\end{align}
\end{subequations}
These equations can be expressed as a canonical Hamiltonian system with Hamiltonian
\begin{equation}
\mf{H}=m\gamma c^2=-p_0,
\end{equation}
and canonical variables $\lr{\bol{p},\bol{x}}$.
Notice that the role played by the geodesic Hamiltonian \eqref{H} is now replaced by minus the 
canonical momentum $p_0$. 
 
The Maxwell-J\"uttner distribution $f_{\rm MJ}$ follows by enforcing the ergodic ansatz 
on the invariant measure defined by the canonical equations \eqref{SR2}.
In particular, following the same line of argument of the previous section, 
the equilibrium distribution function is
\begin{equation}
f_{\rm MJ}=\frac{1}{Z}e^{-\beta m\gamma c^2}=\frac{1}{Z}\exp\left\{-\beta mc^2\sqrt{1+\frac{\bol{p}^2}{m^2c^2}}\right\}.\label{feqMJ}
\end{equation}
This distribution is different from what one obtains by directly substituting the
Minkowski metric tensor $\eta$ into \eqref{feq},
\begin{equation}
f=\frac{1}{Z}\exp\left\{-\frac{\beta}{2m}\lr{-\frac{p_0^2}{c^2}+\bol{p}^2}\right\}.\label{feqSR}
\end{equation}
This is because there exists a fundamental difference between
the derivation of the distribution function $f$ of \eqref{feq} 
and the Maxwell-J\"uttner distribution $f_{\rm MJ}$ above. 
Indeed, while \eqref{feq} is obtained by enforcing ergodicity with respect to proper time $\tau$, 
\eqref{feqMJ} is constructed with $t$ as time variable for the underlying Hamiltonian system.
Hence, in the former case the notion of thermodynamic equilibrium mathematically corresponds to the existence
of a proper time $\tau\rightarrow\infty$ beyond which the probability of finding a particle in
a certain region of the phase space is independent of $\tau$, $\p f/\p\tau=0$. 
However, in the latter case thermodynamic equilibrium is anchored to a particular choice of the time variable, $t$.
The applicability of the distribution functions \eqref{feq} and \eqref{feqMJ} therefore
depends on the validity of the corresponding ergodic assumptions for the underlying dynamical systems.  
Nevertheless, we remark that the choice of proper time $\tau$ does not suffer the ambiguity that occurs in the choice of $t$
and therefore the resulting distribution function $f$ is expected to be more fundamental. 

It is clear that both \eqref{feqMJ} and \eqref{feqSR} reduce to the Maxwell-Boltzmann distribution
\begin{equation}
f_{\rm MB}=\frac{1}{Z}\exp\left\{-\frac{\beta}{2}m\bol{v}^2\right\}, 
\end{equation}
 in the classical limit $\bol{v}^2<<c^2$, with $\bol{v}=d\bol{x}/dt$ (the constants $\exp\left\{-\beta m c^2\right\}$ and $\exp\left\{\frac{\beta p_0^2}{2mc^2}\right\}$ 
appearing in  \eqref{feqMJ} and \eqref{feqSR} when taking the limit can be absorbed in the normalization factor $Z$). 
In the following sections we shall also see that, in the classical limit, the equilibrium distribution function
\eqref{feq} correctly reproduces the classical distribution functions of 
mechanical systems where gravitational and non-gravitational forces are present (if non-gravitational forces are present, the geodesic Hamiltonian is replaced by an appropriate generating function including non-gravitational contributions).

\section{Gas Distribution in Schwarzschild Spacetime}
Let $\lr{R,\theta,\phi}$ and $\lr{r,\phi,z}$ denote a spherical coordinate system and a cylindrical 
coordinate system respectively. 
We consider spacetime metrics of the type
\begin{equation}
ds^2=g_{tt}dt^2+g_{RR}dR^2+g_{\theta\theta}d\theta^2+g_{\phi\phi}d\phi^2.\label{ds2}
\end{equation}
Here, we introduced a new notation $\lr{x^0,x^1,x^2,x^3,p_0,p_1,p_2,p_3}=\lr{t,R,\theta,\phi,p_t,p_R,p_\theta,p_\phi}$, $g_{00}=g_{tt}$, $g_{11}=g_{RR}$, $g_{22}=g_{\theta\theta}$, and $g_{33}=g_{\phi\phi}$. In the following, analogous definitions will be used for non-diagonal terms (if present) and the inverse $g^{ij}$. We shall employ both notations, favoring numerical indexes if summations are present, 
and the spherical coordinates notation to better convey physical meaning of expressions.

An example of \eqref{ds2} is Schwarzschild's exterior solution
\begin{equation}
ds^2=-\lr{1-\frac{R_s}{R}}c^2dt^2+\lr{1-\frac{R_s}{R}}^{-1}dR^2+R^2\lr{d\theta^2+\sin^2\theta d\phi^2}.\label{ds2Sc}
\end{equation}
Here, $R_s=2GM/c^2$ is the Schwarzschild radius, $M$ the central mass responsible for the metric \eqref{ds2Sc}, 
and $G$ the gravitational constant. 

Our aim in this section is to determine the equilibrium spatial distribution of an ensemble of massive particles 
in the presence of a spacetime metric of the form \eqref{ds2}. 
This could be the case of a neutral gas
relaxing within the gravitational field of a central object (e.g. a black hole) with mass $M$ 
shaping spacetime according to \eqref{ds2Sc}. 
Depending on the properties of the relaxation mechanism and the external forces acting on the system, 
certain macroscopic observables may be preserved while the system approaches equilibrium. 
For example, if a classical ensemble of particles interact through elastic collisions, 
total energy and momentum remain constant. 
Or, if interactions possess a symmetry,  
quantities like the total angular momentum will be preserved and
they will eventually affect the equilibrium state of the system. 
To determine candidate macroscopic constraints, 
define the quantities
\begin{equation}
l_z=mg_{\phi\phi}\dot{\phi}=p_{\phi},\label{lz}
\end{equation}
and
\begin{equation}
\bol{l}^2=m^2\lr{g_{\theta\theta}^2\dot{\theta}^2+g_{\theta\theta}g_{\phi\phi}\dot{\phi}^2}=p_{\theta}^2+\frac{g_{\theta\theta}}{g_{\phi\phi}}p_{\phi}^2.\label{l2}
\end{equation}
As usual, the dot denotes differentiation with respect to proper time. 
In the following, we demand that $g_{\theta\theta}d\theta^2+g_{\phi\phi}d\phi^2=R^2\lr{d\theta^2+\sin^2\theta\, d\phi^2}$,  
and that $g_{tt}$ and $g_{RR}$ are radial functions. Then, both $l_{z}$ and $\bol{l}^2$ are constants of motion 
of the geodesic equations \eqref{geo}. 
Classically, $l_z$ is the $z$-component of angular momentum, while
$\bol{l}^2$ its squared modulus. 
We consider a scenario in which forces acting on the system 
(e.g. the gravitational pull of the central mass or collisions among particles) 
do not break the corresponding macroscopic conservation laws, in the sense that
the total $z$-component of the angular momentum $L_z$, and the total angular momentum $\bol{L}^2$ are preserved while the system approaches thermodynamic equilibrium. 
The quantities $L_z$ and $\bol{L}^2$ are defined as
\begin{equation}
L_z=\int_{\Pi}fl_z d^3p d^3x,
\end{equation}
and 
\begin{equation}
\bol{L}^2=\int_{\Pi}f\bol{l}^2d^3p d^3x.
\end{equation}
In the classical setting, 
the radial gravitational force exerted by a spherical central mass 
does not apply torque on the system, resulting in 
conservation of total angular momentum $\bol{L}^2$ and its components.

Consider the case in which particles are mainly following rotational orbits around an axis, say the $z$-axis, 
on the plane $z=0$. 
Let 
$v_{\phi}=p_{\phi}/mr=r\dot{\phi}$ be the velocity of rotation around the $z$-axis. 
Since radial central forces do not apply a net force in the azimuthal
direction, if particles encounters can be approximated by collisions in a classical regime,    
one may assume that 
the system preserves the total azimuthal momentum
\begin{equation}
J_{\phi}=m\int_{\Pi}fv_{\phi}d^3p d^3x.
\end{equation} 
This quantity may be relevant in the description of disk like distributions
arising as a consequence of an initial macroscopic rotation around the $z$-axis. 
 
In the following, constraints like $L_z$, $\bol{L}^2$, and $J_{\phi}$  
will be enforced through the method of Lagrange multipliers
in the variational principle extremizing the entropy of the system.
Conversely, notice that the breaking of a constraint can always be restored by setting
the corresponding Lagrange multiplier to zero. 

\begin{figure}[h]
\hspace*{-0.2cm}\centering
\includegraphics[scale=0.22]{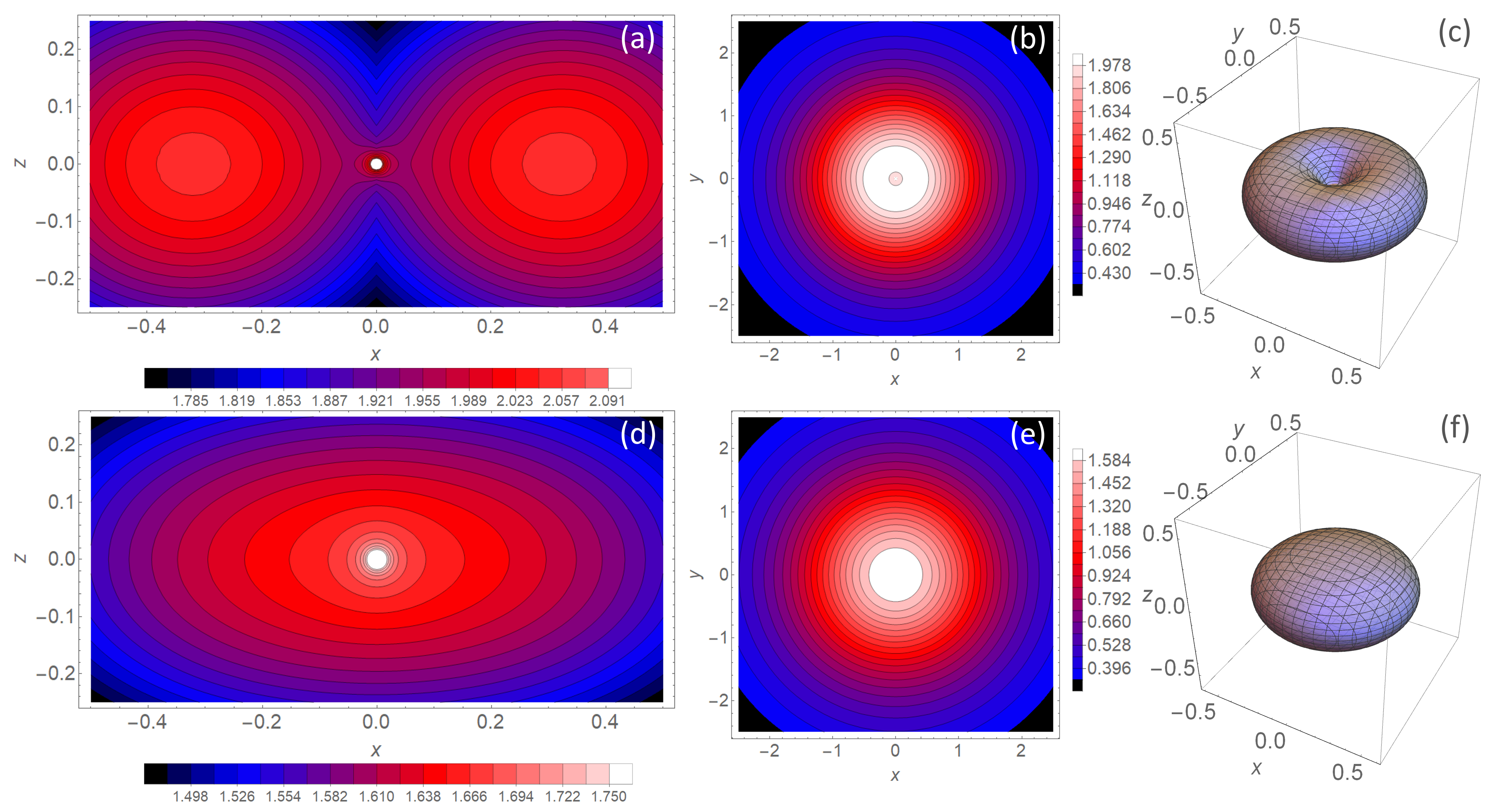}
\caption{\footnotesize Contour plots of the Schwarzschild laboratory density \eqref{rhosclab}
for $m=1$, $c=1$, $R_s=0.001$, $p_t=-mc^2$, $\beta=1$, and $\epsilon=1/2$. (a) Contour plot in the $\lr{x,z}$ plane for $\zeta=1.025$ and $\eta=1/2$. (b) 
Contour plot in the $\lr{x,y}$ plane for $\zeta=1.025$ and $\eta=1/2$. (c) Contour plot in $\lr{x,y,z}$ space for $\zeta=1.025$ and $\eta=1/2$. (d) Contour plot in the $\lr{x,z}$ plane for $\zeta=1.275$ and $\eta=0$. (e) Contour plot in the $\lr{x,y}$ plane for $\zeta=1.275$ and $\eta=0$. (f) Contour plot in $\lr{x,y,z}$ space for $\zeta=1.275$ and $\eta=0$. }
\label{fig1}
\end{figure}

Next, recall that the invariant (Liouville) measure of the system is given by
the phase space volume element
\begin{equation}
d\Pi=d^3p d^3x=dp_{R}dp_{\theta}dp_{\phi}dRd\theta d\phi.\label{vol}
\end{equation}
Let $f$ denote the particle distribution function defined with respect to
the canonical set $\lr{p_{R},p_{\theta},p_{\phi},R,\theta,\phi}$.
The entropy $S$ of the system is then given by Shannon's information measure \eqref{S}.
Then, the equilibrium distribution function is calculated according to the variational principle
\begin{equation}
\delta\lr{S-\alpha N-\beta E-\epsilon \bol{L}^2-\zeta L_z-\eta J_{\phi}}=0,
\end{equation}
where variations are carried out with respect to $f$. 
Here, $\alpha$, $\beta$, $\epsilon$, $\zeta$, and $\eta$ are
Lagrange multipliers, and the total particle number $N$ 
and the total energy $E$ are defined as in \eqref{NE}.  
One obtains
\begin{equation}
f=\frac{1}{Z}\exp\left\{-\beta \mc{H}-\epsilon \bol{l}^2-\zeta l_z-\eta mv_{\phi}\right\},\label{f1}
\end{equation}
with $Z=e^{\lr{1+\alpha}}$. Explicitly, equation \eqref{f1} can be
written as
\begin{equation}
f=\frac{1}{Z}\exp\left\{-\left[\frac{\beta}{2m}\lr{g^{tt}p_t^2+g^{RR}p^2_R}+\lr{\frac{\beta}{2m} g^{\theta\theta}+\epsilon}p^2_{\theta}+\lr{\frac{\beta}{2m} g^{\phi\phi}+\epsilon\frac{R^2}{r^2}}p^2_{\phi}+\lr{\zeta+\eta g^{\phi\phi}r}p_{\phi}\right]\right\}.\label{f2}
\end{equation}
The particle density seen in the spherical coordinate system $\lr{R,\theta,\phi}$ 
is therefore:
\begin{equation}
\rho=\int_{\mathbb{R}^3}fd^3p=\frac{1}{Z}\exp\left\{-\frac{\beta}{2m}g^{tt}p_t^2+\frac{\lr{\zeta+\eta g^{\phi\phi}r}^2}{4\lr{\frac{\beta}{2m} g^{\phi\phi}+\epsilon\frac{R^2}{r^2}}}\right\}\sqrt{\frac{2\pi m}{\beta g^{RR}}}\sqrt{\frac{\pi}{\frac{\beta}{2m} g^{\theta\theta}+\epsilon}}\sqrt{\frac{\pi}{\frac{\beta}{2m} g^{\phi\phi}+\epsilon\frac{R^2}{r^2}}}.\label{rho_gen}
\end{equation}
Recall that $dxdydz=RrdRd\theta d\phi$. Hence, the particle density $\rho_{\rm lab}$ in the
laboratory frame $\lr{x,y,z}$ is given by
\begin{equation}
\rho_{\rm lab}=\frac{\rho}{Rr}=
\frac{1}{ZRr}\exp\left\{-\frac{\beta}{2m}g^{tt}p_t^2+\frac{\lr{\zeta+\eta g^{\phi\phi}r}^2}{4\lr{\frac{\beta}{2m} g^{\phi\phi}+\epsilon\frac{R^2}{r^2}}}\right\}\sqrt{\frac{2\pi m}{\beta g^{RR}}}\sqrt{\frac{\pi}{\frac{\beta}{2m} g^{\theta\theta}+\epsilon}}\sqrt{\frac{\pi}{\frac{\beta}{2m} g^{\phi\phi}+\epsilon\frac{R^2}{r^2}}}.
\end{equation}
For the Schwarzschild metric $g^{tt}=-(1-R_s/R)^{-1}c^{-2}$, $g^{RR}=1-R_s/R$, $g^{\theta\theta}=1/R^2$, and $g^{\phi\phi}=1/r^2$. Hence, we obtain the Schwarzschild laboratory density
distribution
\begin{equation}
\rho_{\rm lab}^{\rm Sc}=\frac{\pi}{Z}\sqrt{\frac{2\pi m}{\beta}}
\exp\left\{\frac{\beta}{2mc^2}\frac{p^2_t}{1-\frac{R_s}{R}}+\frac{\lr{r\zeta+\eta}^2}{4\lr{\frac{\beta}{2m} +\epsilon R^2}}\right\}
\frac{1}{\sqrt{1-\frac{R_s}{R}}}\frac{1}{\frac{\beta}{2m}+\epsilon R^2}
.\label{rhosclab}
\end{equation}
Notice that this density is well defined only outside the Schwarzschild radius $R_s$.
Assuming $\beta, m > 0$ and $\epsilon\geq 0$, we have 
\begin{subequations}
\begin{align}
\lim_{R\rightarrow R_s^+}\rho_{\rm lab}^{\rm Sc}=&+\infty,\\
\lim_{r\rightarrow \infty}\rho_{\rm lab}^{\rm Sc}=&\begin{cases}
+\infty~~~~{\rm if}~~\epsilon=0,~~\zeta\neq 0\\
0~~~~{\rm if}~~\epsilon>0,~~\zeta= 0\\
\frac{1}{Z}\lr{\frac{2\pi m}{\beta}}^{3/2}\exp\left\{\frac{\beta p^2_t}{2mc^2}+\frac{m\eta^2}{2\beta}\right\}~~~~{\rm if}~~\epsilon=0,~~\zeta= 0\\
0~~~~{\rm if}~~\epsilon>0,~~\zeta\neq0
\end{cases}.
\end{align}
\end{subequations}
Figure \ref{fig1} shows the profile of the Schwarzschild laboratory density distribution \eqref{rhosclab} 
for specific choices of physical parameters and Lagrange multipliers. 
The contours of \eqref{rhosclab} are essentially discoidal, although the Lagrange multiplier $\eta$ associated with $J_{\phi}$ 
introduces a central distortion that produces loboidal structures resembling a dipole field.    

The classical mass density distribution can be recovered by
considering the limit $R_s/R<<1$, $\beta mc^2>>1$ while recalling that $p_t=-mc^2$ 
for an ensemble initially at rest in Minkowski spacetime. 
Physically, these conditions can be respectively understood as follows: 
1) particles are far from the event horizon, and therefore the central mass  
affects them through a classical gravitational potential.
2) thermal fluctuations $\tilde{\bol{v}}$, which scale as $\beta^{-1}\sim m\tilde{\bol{v}}^2/2$, 
are negligible when compared with the rest energy of each particle. 
We observe that the assumption $p_t=-mc^2$ can also be interpreted as a result of the limit $p_t\rightarrow -mc^2$ 
occurring when particle velocities are small compared with the speed of light, so that 
the Lorentz factor satisfies $\gamma\rightarrow 1$.
At first order in $R_s/R$, we have
\begin{equation}
\rho_{\rm lab}^{\rm Sc}=\frac{\pi}{Z}\sqrt{\frac{2\pi m}{\beta}}
\frac{1}{\frac{\beta}{2m}+\epsilon R^2}
\exp\left\{\frac{\beta p^2_t}{2mc^2}+\frac{GM}{Rc^2}\lr{1+\frac{\beta p^2_t}{mc^2}}+\frac{\lr{r\zeta+\eta}^2}{4\lr{\frac{\beta}{2m} +\epsilon R^2}}\right\}.\label{rhosclab1}
\end{equation}
Here, we used the expression for the Schwarzschild radius $R_s=2GM/c^2$. 
Then, taking the limit $p_t\rightarrow mc^2$ gives
\begin{equation}
\rho_{\rm lab}^{\rm Sc}=\frac{\pi}{Z}\sqrt{\frac{2\pi m}{\beta}}
\frac{1}{\frac{\beta}{2m}+\epsilon R^2}
\exp\left\{\beta mc^2\left[\frac{1}{2}+\frac{GM}{Rc^2}\lr{\frac{1}{\beta m c^2}+1}+\frac{1}{\beta mc^2}\frac{\lr{r\zeta+\eta}^2}{4\lr{\frac{\beta}{2m} +\epsilon R^2}}\right]\right\}.\label{rhosclab2}
\end{equation}
Notice that equation \eqref{rhosclab2} predicts an effective gravitational constant
\begin{equation}
G'=G\lr{1+\frac{1}{\beta mc^2}}.\label{Gprime}
\end{equation} 
Since in the classical limit $\beta^{-1}$ represents the inverse of the thermodynamic temperature of the system in units of energy, 
equation \eqref{Gprime} implies that a finite temperature increases the effective
gravitational force exchanged by interacting particles. 
Mathematically, the thermodynamic correction arising in \eqref{Gprime} is a consequence of the
term involving $p_t=p_0$ in the geodesic Hamiltonian (i.e. the `kinetic' energy associated with the speed of the 
individual clocks $\dot{t}$). 
In most cases the correction appears to be negligible. For example, 
in the case of an electron gas one obtains  
\begin{equation}
\frac{1}{\beta m_e c^2}=\frac{k_BT_e}{m_e c^2}\sim 1.7\, 10^{-10}\,K^{-1}\, T_e.\label{TeCorr}
\end{equation}  
Here, $k_B$ is the Boltzmann constant, $T_e$ the electron temperature, $m_e$ the electron mass, and $K$ the Kelvin unit. 
The value \eqref{TeCorr} is small even for an electron temperature of the order of a billion Kelvin. 
Taking the limit $\beta mc^2\rightarrow +\infty$, one obtains
\begin{equation}
\rho_{\rm lab}^{\rm Sc}=\frac{\pi}{Z}\sqrt{\frac{2\pi m}{\beta}}
\frac{1}{\frac{\beta}{2m}+\epsilon R^2}
\exp\left\{\beta\lr{\frac{ mc^2}{2}+\frac{GMm}{R}}\right\}.\label{rhosclab3}
\end{equation} 
The second term in the exponential is the classical gravitational potential energy
normalized by the temperature $\beta^{-1}$. 
Finally, breaking conservation of total angular momentum $\bol{L}^2$ by setting $\epsilon=0$
and redefining the normalization constant as $Z'=Z\exp\left\{-\beta \frac{mc^2}{2}\right\}$ 
one arrives at the classical density distribution
\begin{equation}
\rho_{\rm lab}^{\rm Sc}=\frac{1}{Z'}\lr{\frac{2\pi m}{\beta}}^{3/2}
\exp\left\{\beta\lr{\frac{GMm}{R}}\right\}.\label{rhosclab4}
\end{equation}

In addition to the spatial density distribution, let us consider how the
typical velocity of rotation around the $z$-axis is affected by spacetime curvature. 
This physical observable may be relevant, for example, in the study of the  
speed of mass distributions rotating around a galacting center.  
At equilibrium and at a given point in spacetime, this velocity can be calculated as
\begin{equation}
\begin{split}
v_{\rm rot}&=\frac{1}{\rho}\int_{\mathbb{R}^3}fv_{\phi}d^3p=\\
&\frac{rg^{\phi\phi}}{mZ\rho}\int_{\mathbb{R}^3}\exp\left\{
-\left[\frac{\beta}{2m}\lr{g^{tt}p_t^2+g^{RR}p^2_R}+\lr{\frac{\beta}{2m} g^{\theta\theta}+\epsilon}p^2_{\theta}\right]\right\}\\
&\exp\left\{-\left[\lr{\frac{\beta}{2m} g^{\phi\phi}+\epsilon\frac{R^2}{r^2}}p^2_{\phi}+\lr{\zeta+\eta g^{\phi\phi}r}p_{\phi}\right]
\right\}p_{\phi}d^3p\\
=&-\frac{rg^{\phi\phi}}{2mZ\rho}\exp\left\{-\frac{\beta}{2m}g^{tt}p_t^2+\frac{\lr{\zeta+\eta g^{\phi\phi}r}^2}{4\lr{\frac{\beta}{2m} g^{\phi\phi}+\epsilon\frac{R^2}{r^2}}}\right\}\sqrt{\frac{2\pi m}{\beta g^{RR}}}\sqrt{\frac{\pi}{\frac{\beta}{2m} g^{\theta\theta}+\epsilon}}\frac{\sqrt{\pi}\lr{\zeta+\eta g^{\phi\phi}r}}{\lr{\frac{\beta}{2m} g^{\phi\phi}+\epsilon\frac{R^2}{r^2}}^{3/2}}\\
=&-\frac{rg^{\phi\phi}}{2m}\frac{\zeta+\eta g^{\phi\phi} r}{\frac{\beta}{2m} g^{\phi\phi}+\epsilon\frac{R^2}{r^2}}.
\end{split}
\end{equation}
Here, equation \eqref{rho_gen} was used. 
Substituting the coefficients of the Schwarzschild metric one obtains
\begin{equation}
v_{\rm rot}^{\rm Sc}=-\frac{r\zeta+\eta}{\beta+2m\epsilon R^2}.\label{vrotSc}
\end{equation}
Assuming $\beta, m > 0$ and $\epsilon\geq 0$, we have 
\begin{equation}
\lim_{r\rightarrow\infty}v_{\rm rot}^{\rm Sc}=
\begin{cases}
-\infty ~~~~{\rm if}~~\epsilon=0,~~\zeta> 0\\
+\infty ~~~~{\rm if}~~\epsilon=0,~~\zeta< 0\\
0~~~~{\rm if}~~\epsilon>0,~~\zeta= 0\\
-\frac{\eta}{\beta}~~~~{\rm if}~~\epsilon=0,~~\zeta=0\\
0~~~~{\rm if}~~\epsilon>0,~~\zeta\neq 0
\end{cases}.
\end{equation}
Recall that setting a Lagrange multiplier to zero is equivalent to
breaking the corresponding constraint. 
When $\epsilon=0$, $\zeta\neq 0$ 
both density and rotation velocity diverge at large radii. 
The case $\epsilon> 0$, $\zeta=0$ leads to decreasing
density and rotation velocity, which eventually scale as $\rho_{\rm lab}^{\rm Sc}\sim R^{-2}$ and $v_{\rm rot}\sim R^{-2}$. 
The case $\epsilon=\zeta=0$ gives a decreasing density profile, which is a function of the ratio $R_s/R$, 
and a constant rotation speed at all points in spacetime, 
$v_{\rm rot}^{\rm Sc}=-\eta/\beta$.
It should be emphasized that this configuration is a result of
the general relativistic distortion of the Minkowski metric
(the factors depending on $1/(1-R_s/R)$ in \eqref{rhosclab}) combined
with conservation of the linear momentum $J_{\phi}$.
Finally, in the case $\epsilon>0$, $\zeta\neq 0$, 
density and rotation velocity decrease at large radii, 
$\rho_{\rm lab}^{\rm Sc}\sim R^{-2}$, $v_{\rm rot}^{\rm Sc}\sim rR^{-2}$.

\begin{figure}[h]
\hspace*{-0cm}\centering
\includegraphics[scale=0.3]{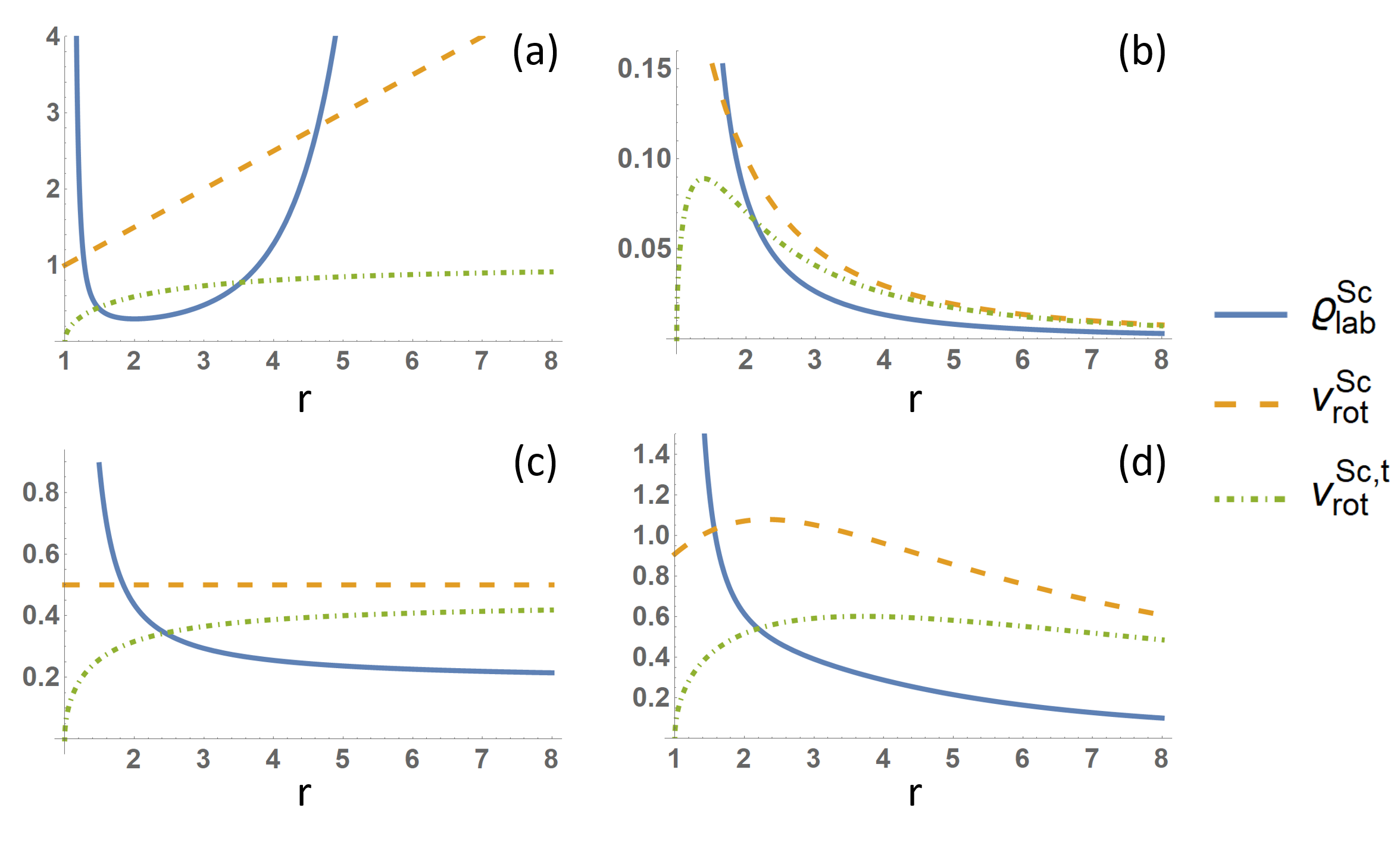}
\caption{\footnotesize Radial profiles of Schwarzschild laboratory density \eqref{rhosclab}, 
average rotation velocity in proper time \eqref{vrotSc}, and average rotation velocity in time $t$ \eqref{vrotSct} on the plane $z=0$ for $m=1$, $c=1$, $R_s=1$, $p_t=-mc^2$, $\beta=1$, and $\eta=-1/2$. Density 
is given in arbitrary units. 
(a) The case $\epsilon=0$, $\zeta=-1/2$. (b) The case $\epsilon=1/2$, $\zeta=0$. (c) The case $\epsilon=0$, $\zeta=0$. (d) The case $\epsilon=0.05$, $\zeta=-1/2$. Notice that all densities diverge in correspondence of the Schwarzschild radius $R_s$.}
\label{fig2}
\end{figure}

The rotation velocity \eqref{vrotSc} is defined with respect to proper time $\tau$.
However, for a stationary observer in the Cartesian coordinate frame $\lr{t,x,y,z}$,  
the measured average rotation velocity is distorted according to
\begin{equation}
v_{\rm rot}^{\rm Sc, t}=v_{\rm rot}^{\rm Sc}\frac{d\tau}{dt}=v_{\rm rot}^{\rm Sc}\sqrt{\frac{1-\frac{R_s}{R}}{1+\lr{\frac{v_{\rm rot}^{Sc}}{c}}^2}}=-\frac{r\zeta+\eta}{\beta+2m\epsilon R^2}\sqrt{\frac{1-\frac{R_s}{R}}{1+\frac{\lr{r\zeta+\eta}^2}{c^2\lr{\beta+2m\epsilon R^2}^2}}}.\label{vrotSct}
\end{equation}
Here, we used the fact that for an object rotating in the $\lr{x,y}$ plane
the Schwarzschild metric leads to the following relationship between 
proper time and coordinate time, $c^2d\tau^2=c^2(1-R_s/R)dt^2-\lr{v_{\rm rot}^{Sc}}^2d\tau^2$  
(observe that integrals of the type $\int_{\mathbb{R}^3}fp_Rd^3p$ and $\int_{\mathbb{R}^3}fp_\theta d^3p$ vanish, implying that on average $\dot{R}=\dot{\theta}=0$ and particles simply rotate in the $\lr{x,y}$ plane).  
From \eqref{vrotSct} we thus have 
\begin{equation}
\lim_{R\rightarrow R_s^+}v_{\rm rot}^{\rm Sc, t}=0,
\end{equation}
and also
\begin{equation}
\lim_{r\rightarrow\infty}v_{\rm rot}^{\rm Sc, t}=
\begin{cases}
-c~~~~{\rm if}~~\epsilon=0,~~\zeta> 0\\
c~~~~{\rm if}~~\epsilon=0,~~\zeta< 0\\
0~~~~{\rm if}~~\epsilon>0,~~\zeta= 0\\
-\frac{\eta}{\beta}\frac{1}{\sqrt{1+\lr{\frac{\eta}{\beta c}}^2}}~~~~{\rm if}~~\epsilon=0,~~\zeta=0\\
0~~~~{\rm if}~~\epsilon>0,~~\zeta\neq 0
\end{cases}.
\end{equation}
Figure \ref{fig2} shows radial profiles of the Schwarzschild laboratory density \eqref{rhosclab}, 
average rotation velocity in proper time \eqref{vrotSc}, and average rotation velocity in time $t$ \eqref{vrotSct} on the plane $z=0$ 
for different values of the Lagrange multipliers $\epsilon$ and $\zeta$, which express conservation of angular momentum. 
Physical units are chosen so that particle mass $m$, speed of light $c$, Schwarzschild radius $R_s$, and inverse temperature $\beta$ 
are unity. Notice that certain configurations, such as (c) and (d) in figure \ref{fig2} are compatible
with a decreasing density distribution and a constant or increasing average rotation velocity in time $t$ at radii $r>R_s$. 
 
\begin{figure}[h]
\hspace*{-0cm}\centering
\includegraphics[scale=0.26]{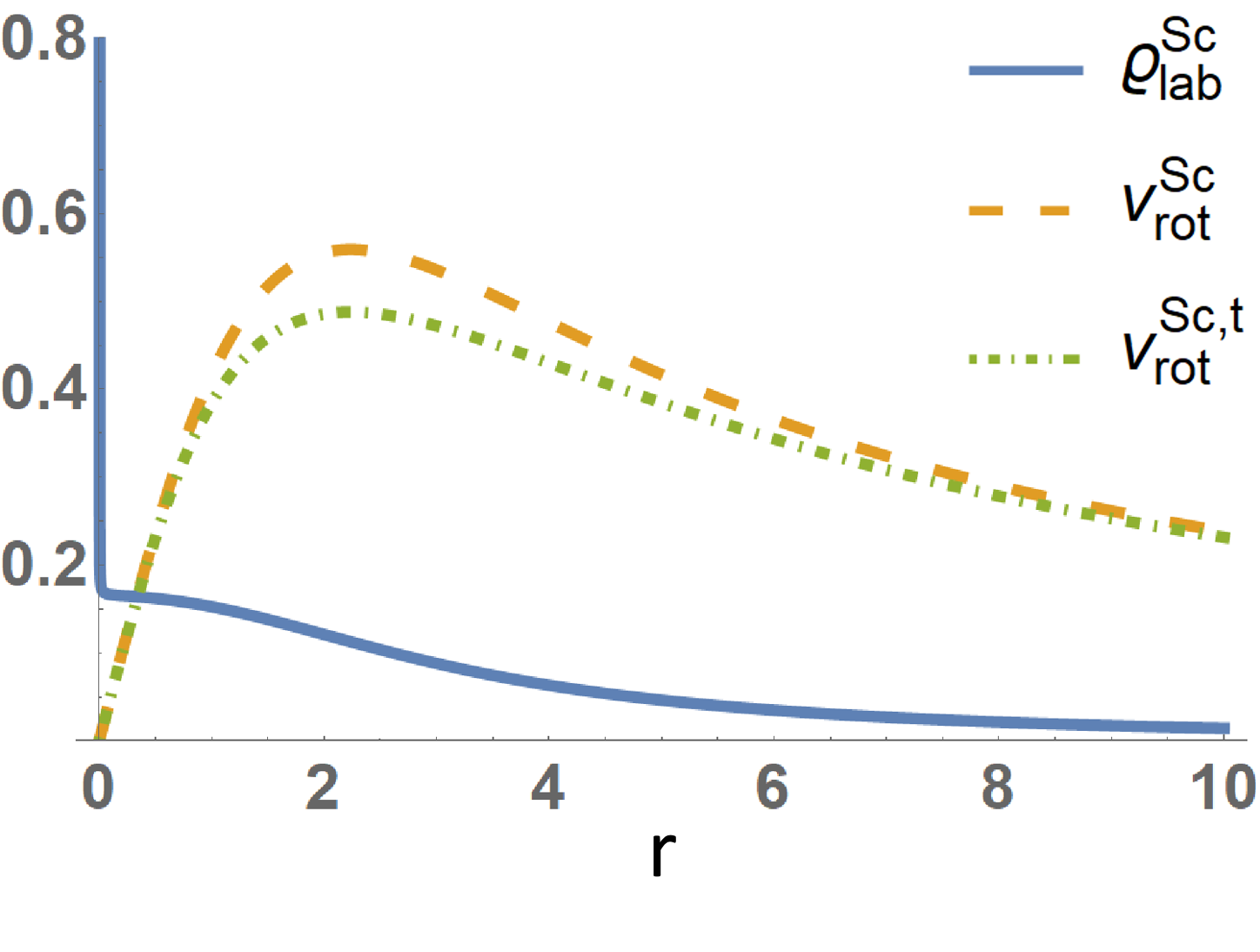}
\caption{\footnotesize Radial profiles of Schwarzschild laboratory density \eqref{rhosclab}, 
average rotation velocity in proper time \eqref{vrotSc}, and average rotation velocity in time $t$ \eqref{vrotSct} on the plane $z=0$ for $m=1$, $c=1$, $R_s=10^{-3}$, $p_t=-mc^2$, $\beta=1$, $\epsilon=0.1$, $\zeta=-1/2$, and $\eta=0$. Density is given in arbitrary units. Notice that density diverges at the Schwarzschild radius $R_s$.}
\label{fig3}
\end{figure}

It is useful to study how the position of the radial peak in average rotation speed $v_{\rm rot}^{\rm Sc,t}$ depends 
on the Schwarzschild radius $R_s$ and the Lagrange multipliers $\epsilon$ and $\zeta$. 
To further simplify the calculations, let us assume that $\eta=0$. 
Setting $m=c=\beta=1$ and $z=0$, expression \eqref{vrotSct} becomes
\begin{equation}
v_{\rm rot}^{\rm Sc,t}=-\frac{r\zeta}{1+2\epsilon r^2}\sqrt{\frac{1-\frac{R_s}{r}}{1+\frac{r^2\zeta^2}{\lr{1+2\epsilon r^2}^2}}}.
\end{equation}
The extrema of this function can be evaluated by setting $dv_{\rm rot}^{\rm Sc,t}/dr=0$. One obtains
\begin{equation}
2r\lr{-1+4\epsilon^2r^4}-R_s\lr{-1+4\epsilon r^2+12\epsilon^2r^4+\zeta^2r^2}=0.
\end{equation}
Denoting with $r_{\ast}$ a solution of the equation above, 
\begin{equation}
R_s=\frac{2\lr{-1+4\epsilon^2 r_{\ast}^4}r_{\ast}}{-1+\lr{4\epsilon+\zeta^2}r_{\ast}^2+12\epsilon^4r_{\ast}^4}.
\end{equation}
Hence, the Lagrange multipliers $\epsilon$ and $\zeta$ introduce nonlinearity in the relationship
between the Schwarzschild radius $R_s$ and the radial peak $r_{\ast}$ of $v_{\rm rot}^{\rm Sc,t}$ 
. 
This nonlinearity makes it possible to achieve configurations in which  
the position of the maximum is faraway from the Schwarzschild radius of the central mass, $R_s/r_{\ast}<<1$.
For example, setting $R_s=10^{-3}$, $\epsilon=0.1$, and $\zeta=-1/2$, gives $r_{\ast}\sim 2.237$.
Figure \ref{fig3} shows the corresponding radial profiles of $\rho_{\rm lab}^{\rm Sc}$, $v_{\rm rot}^{Sc}$, and $v_{\rm rot}^{Sc,t}$ 
on the plane $z=0$. 

\section{Gas Distribution in Kerr-Newman Spacetime}

\begin{figure}[h]
\hspace*{-0cm}\centering
\includegraphics[scale=0.24]{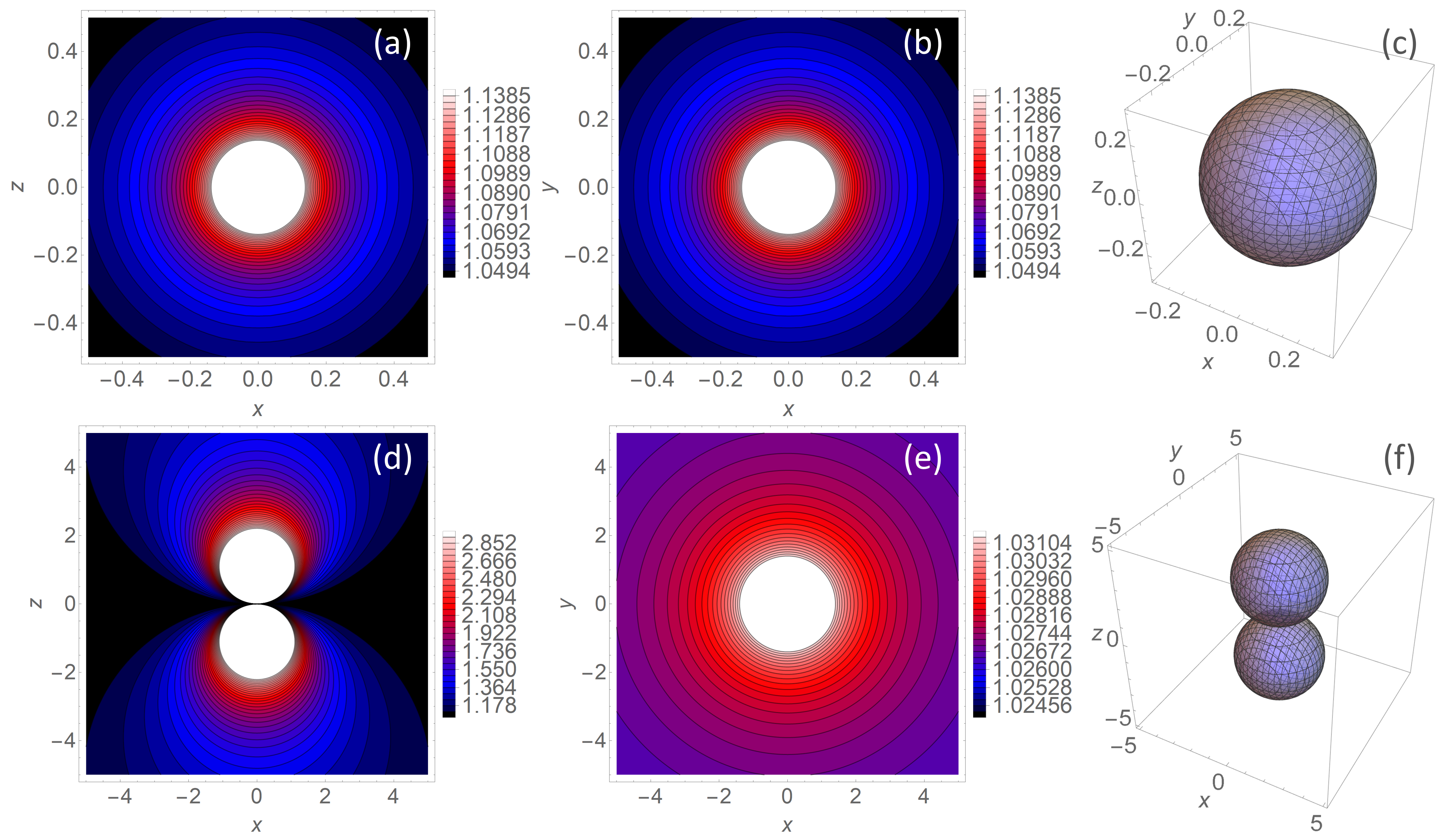}
\caption{\footnotesize Contour plots of the laboratory density distribution \eqref{rhoKNlab} for $m=0.05$, $c=1$, $p_t=-mc^2$, $\beta=1$, $e=1$, $G=1$, and $\epsilon_0=1$. (a) Contour plot in the $\lr{x,z}$ plane for $R_s=0.02$, $a=0.001$ and $R_Q=0.009$. (b) Contour plot in the $\lr{x,y}$ plane for $R_s=0.02$, $a=0.001$ and $R_Q=0.009$. (c) Contour plot in $\lr{x,y,z}$ space for $R_s=0.02$, $a=0.001$ and $R_Q=0.009$. (d) Contour plot in the $\lr{x,z}$ plane for $R_s=0.01$, $a=3$ and $R_Q=0.01$. (e) Contour plot in the $\lr{x,y}$ plane for $R_s=0.01$, $a=3$ and $R_Q=0.01$. (f) Contour plot in $\lr{x,y,z}$ space for $R_s=0.01$, $a=3$ and $R_Q=0.01$. Notice that the configuration of (a), (b), and (c) satisfies $R_s^2>4\lr{a^2+R_q^2}$, while that of (d), (e), and (f) satisfies $R_s^2<4\lr{a^2+R_q^2}$.}
\label{fig4}
\end{figure}

In this section we consider the Kerr-Newman metric
\begin{equation}
\begin{split}
ds^2=&g_{tt}dt^2+2g_{t\phi}dtd\phi+g_{RR}dR^2+g_{\theta\theta}d\theta^2+g_{\phi\phi}d\phi^2\\
=&-\frac{c^2}{q^2}\lr{\Delta-a^2\sin^2\theta}dt^2+\frac{2ac\sin^2\theta}{q^2}\lr{R^2_Q-R_sR}dtd\phi
+\frac{q^2}{\Delta}dR^2+q^2d\theta^2\\&+\frac{\sin^2\theta}{q^2}\left[
\lr{R^2+a^2}^2-a^2\Delta \sin^2\theta
\right]d\phi^2.\label{ds2KN}
\end{split}
\end{equation} 
Here 
$a=J/Mc$, $R_Q^2=Q^2G/4\pi\epsilon_0c^4$, with $J$ and $Q$ the angular momentum and the electric charge of the central mass respectively, $\epsilon_0$ the vacuum permittivity, and we defined
\begin{subequations}
\begin{align}
\Delta=&R^2-R_sR+a^2+R_Q^2,\\
q^2=&R^2+a^2\cos^2\theta.
\end{align}
\end{subequations}
Notice that equation \eqref{ds2KN} reduces to the Schwarzschild metric when $a=R_Q=0$. 
We also recall that \eqref{ds2KN} corresponds to spacetime around a black hole  
when $R_s^2\geq 4(a^2+R_Q^2)$. In such case, singularities (event horizons) occur at 
\begin{equation}
R_{\pm}=\frac{R_s}{2}\pm\sqrt{\frac{R_s^2}{4}-a^2-R^2_Q}.
\end{equation} 
Due to the electromagnetic field, charged particle orbits are not
pure geodesics, but they are affected by the Lorentz force. 
Setting $\lr{x^0,x^1,x^2,x^3}=\lr{t,R,\theta,\phi}$, the equations of motion are:
\begin{equation}
\ddot{x}^i=-\Gamma^i_{jk}\dot{x}^j\dot{x}^k+\frac{e}{m}F^{ik}\dot{x}^jg_{jk},\label{geqem}
\end{equation} 
where $F^{ij}=g^{i\alpha}g^{i\beta}\lr{\p_\alpha A_\beta-\p_\beta A_\alpha}$ is the contravariant Maxwell-Faraday tensor
associated with the four-potential $A_i$, $i=0,1,2,3$, and $e$ the particle electric charge. 
The Hamiltonian associated with system \eqref{geqem} is
\begin{equation}
H=\frac{1}{2m}g^{ij}\lr{p_i-eA_i}\lr{p_j-eA_j}.\label{Hgeqem}
\end{equation} 
Here, canonical variables are given by $\lr{p_i,x^i}$, $i=0,...,3$, and the four potential $A=A_idx^i$ has expression
\begin{equation}
A=-\frac{Q}{4\pi \epsilon_0c}\frac{R}{q^2}\lr{cdt-a \sin^2\theta d\phi}.
\end{equation}
Now observe that, since $\p g_{ij}/\p t=0$, the Hamiltonian is symmetric in time $\p H/\p t=0$, and the canonical momentum $p_t$ is again a constant, while the variable $t$ evolves independently according to $\dot{t}=\p H/\p p_t$. 
Therefore, we can apply the construction of section 2 
over a level set of $p_t$ embedded in the original phase space, 
and derive the equilibrium distribution function $f$ of equation \eqref{feq} associated with the reduced system with Hamiltonian 
\begin{equation}
\mc{H}=\frac{1}{2m}\left[g^{00}\lr{p_0-eA_0}^2+2g^{0i}\lr{p_0-eA_0}\lr{p_i-eA_i}+\mf{g}^{ij}\lr{p_i-eA_i}\lr{p_j-eA_j}\right], 
\end{equation}
and canonical variables $\lr{p_i,x^i}$, $i=1,2,3$. 
The metric coefficients are also symmetric with respect to $\phi$, implying conservation of the canonical momentum
\begin{equation}
p_\phi=\frac{m\dot{\phi}}{g^{\phi\phi}}-\frac{g^{t\phi}}{g^{\phi\phi}}\lr{p_t-eA_t}+eA_\phi.
\end{equation} 
The system also possesses an additional invariant, the Carter constant, which replaces the quantity $\bol{l}^2$ (equation \eqref{l2})
of the Schwarzschild case.  
One may enforce additional macroscopic constraints on the distribution function $f$, 
such as conservation of the functional 
\begin{equation}
\mc{L}_z=\int_{\Pi}fp_\phi d^3p d^3x,\label{p3KN} 
\end{equation}
through a Lagrange multiplier. 
The results are analogous to the Schwarzschild case. For example, 
following the same steps of the previous section one can verify that
the constraint \eqref{p3KN} leads to an average azimuthal rotation velocity 
of the form \eqref{vrotSc} with $\epsilon=\eta=0$. 
However, in the remainder of this section we shall assume that
no additional constraints are present, so that the equilibrium distribution function has the form $f=Z^{-1}e^{-\beta\mc{H}}$,  
and instead focus on the kinetic energy associated with rotational motion. 
To this end, first we need to 
evaluate the density distribution $\rho=\rho\lr{R,\theta,\phi}$ corresponding to the equilibrium distribution function $f$. Care is needed when handling non-diagonal terms in the metric tensor $g_{ij}$. Define $\xi_t=p_t-eA_t$ and $\xi_\phi=p_\phi-eA_\phi$. We have the Kerr-Newman density distribution
\begin{equation}
\begin{split}
\rho^{\rm KN}=&\frac{1}{Z}\exp\left\{-\frac{\beta}{2m} g^{tt}\xi_t^2\right\}\int_{\mathbb{R}^3}\exp\left\{
-\frac{\beta}{2m}\lr{2g^{t\phi}\xi_t\xi_\phi+g^{RR}p_R^2+g^{\theta\theta}p_\theta^2+g^{\phi\phi}\xi_\phi^2
}\right\}d^3p\\
=&\lr{\frac{2\pi m}{\beta}}^{3/2}\frac{1}{Z\sqrt{g^{RR}g^{\theta\theta}g^{\phi\phi}}}\exp\left\{\frac{\beta}{2m}\left[\frac{\lr{g^{t\phi}}^2}{g^{\phi\phi}}-g^{tt}\right]\xi_t^2\right\}.\label{rhoKN}
\end{split}
\end{equation}
Next, observe that the 4-dimensional Kerr-Newman metric tensor $g_{ij}$ has matrix form
\begin{equation}
\begin{bmatrix}g_{tt}&0&0&g_{t\phi}\\0&g_{RR}&0&0\\0&0&g_{\theta\theta}&0\\g_{t\phi}&0&0&g_{\phi\phi}\end{bmatrix}.
\end{equation} 
It follows that the contravariant metric tensor $g^{ij}$ has matrix form
\begin{equation}
\frac{1}{g_{tt}g_{\phi\phi}-g_{t\phi}^2}\begin{bmatrix}g_{\phi\phi}&0&0&-g_{t\phi}\\0&\lr{g_{tt}g_{\phi\phi}-g_{t\phi}^2}g_{RR}^{-1}&0&0\\0&0&\lr{g_{tt}g_{\phi\phi}-g_{t\phi}^2}g_{\theta\theta}^{-1}&0\\-g_{t\phi}&0&0&g_{tt}\end{bmatrix}.\label{contragKN}
\end{equation}
In the laboratory frame $\lr{x,y,z}$, the density distribution \eqref{rhoKN} therefore transforms to
\begin{equation}
\begin{split}
\rho^{\rm KN}_{\rm lab}=&\lr{\frac{2\pi m}{\beta}}^{3/2}\frac{1}{ZrR}\sqrt{g_{RR}g_{\theta\theta}\lr{g_{\phi\phi}-\frac{g_{t\phi}^2}{g_{tt}}}}\exp\left\{-\frac{\beta \xi_t^2}{2mg_{tt}}\right\}\\
=&\lr{\frac{2\pi m}{\beta}}^{3/2}\frac{q}{ZR^2\sqrt{\Delta}}
\sqrt{\lr{R^2+a^2}^2+a^2\sin^2\theta\left[\frac{\lr{R_Q^2-R_sR}^2}{\Delta-a^2\sin^2\theta}-\Delta\right]}\\
&\exp\left\{\frac{\beta\xi_t^2}{2mc^2}\frac{q^2}{\Delta-a^2\sin^2\theta}\right\}
.\label{rhoKNlab}
\end{split}
\end{equation}
The limit to the Schwarzschild case can be obtained by setting $a=R_Q=0$.
We also have
\begin{equation}
\lim_{r\rightarrow\infty}\rho_{\rm lab}^{\rm KN}=\frac{1}{Z}\lr{\frac{2\pi m}{\beta}}^{3/2}\exp\left\{\frac{\beta p_t^2}{2m c^2}\right\}.
\end{equation}
In figure \ref{fig4}, contours of the Kerr-Newman laboratory density distribution \eqref{rhoKNlab} are shown for given choices of 
physical units and parameters. Notice that level sets of $\rho^{\rm KN}_{\rm lab}$ may form paired bulges that extend along the $z$-axis.
These structures progressively vanishes when the rotation of the central mass, quantified by $a$, approaches zero. 

At a given point in space, we define the rotational kinetic energy 
\begin{equation}
K_{\rm rot}^{\rm KN}=\frac{m}{2\rho^{KN}}\int_{\mathbb{R}^3}fv^{2}_{\phi}d^3p,\label{KKN}
\end{equation}
with $v_{\phi}=r\dot{\phi}$. 
The quantity $K^{\rm KN}_{\rm rot}$ measures the typical energy dedicated by particles to azimuthal rotations, 
and it corresponds to the momentum-space average $\langle~\rangle_{p}$ of squared rotation velocity, 
\begin{equation}
K^{\rm KN}_{\rm rot}=\frac{m}{2}\langle v_{\phi}^2\rangle_{p}.
\end{equation}
Recalling that $m\dot{\phi}=m\p H/\p p_\phi=g^{t\phi}\xi_t+g^{\phi\phi}\xi_\phi$, 
the quantity \eqref{KKN} can be evaluated explicitly as below:
\begin{equation}
\begin{split}
K_{\rm rot}^{\rm KN}=&\frac{r^2}{2mZ\rho^{\rm KN}}\int_{\mathbb{R}^3}\exp\left\{-\frac{\beta}{2m}\left[g^{tt}\xi_t^2+2g^{t\phi}\xi_t\xi_\phi+g^{RR}p_R^2+g^{\theta\theta}p_{\theta}^2+g^{\phi\phi}\xi_{\phi}^2\right]\right\}\lr{g^{t\phi}\xi_t+g^{\phi\phi}\xi_\phi}^2d^3p\\
=&\lr{\frac{2\pi m}{\beta}}^{3/2}\frac{r^2}{2\beta Z\rho^{\rm KN}}{\sqrt{\frac{g^{\phi\phi}}{g^{RR}g^{\theta\theta}}}}\exp\left\{\frac{\beta}{2m}
\left[\frac{\lr{g^{t\phi}}^2}{g^{\phi\phi}}-g^{tt}\right]\xi_t^2\right\}\\
=&\frac{1}{2\beta}r^2g^{\phi\phi}
.\label{KKN3}
\end{split}
\end{equation}
Here, we used equation \eqref{rhoKN}.
In flat spacetime $R_s=R_Q=a=0$ equation \eqref{KKN3} reduces to 
the classical result $K^{\rm KN}_{rot}=1/2\beta$ since in this case $g^{\phi\phi}=1/r^2$.
Next, expressing $g^{\phi\phi}$ in terms of covariant components through \eqref{contragKN}, one obtains
\begin{equation}
K_{\rm rot}^{\rm KN}=\frac{1}{2\beta}\frac{r^2g_{tt}}{g_{tt}g_{\phi\phi}-g_{t\phi}^2}=\frac{1}{2\beta}\frac{R^2q^2}{\lr{R^2+a^2}^2+a^2\sin^2\theta\left[\frac{\lr{R^2_Q-RR_s}^2}{\Delta-a^2\sin^2\theta}-\Delta\right]}.\label{KKN4}
\end{equation}
Assuming $\beta\neq 0$, it follows that
\begin{equation}
\lim_{r\rightarrow \infty}K^{\rm KN}_{\rm rot}=\frac{1}{2\beta}.
\end{equation}
Therefore, the rotational kinetic energy $K_{\rm rot}^{\rm KN}$ 
approaches a constant value at large radii. When $\beta=1/k_BT$, this value corresponds to classical equipartition. 
Finally, for a particle rotating with azimuthal velocity $\sqrt{v_{\phi}^2}=\sqrt{2K_{\rm rot}^{\rm KN}/m}$, 
the rotational kinetic energy observed in the $\lr{t,x,y,z}$ reference frame is
\begin{equation}
\begin{split}
K_{\rm rot}^{\rm KN, t}=&K_{\rm rot}^{\rm KN}\lr{\frac{d\tau}{dt}}^2
=K_{\rm rot}^{\rm KN}\left[\frac{-\frac{g_{t\phi}}{r}\sqrt{\frac{2K^{\rm KN}_{\rm rot}}{m}}
+\sqrt{\frac{g_{t\phi}^2}{r^2}\frac{2K^{\rm KN}_{\rm rot}}{m}-g_{tt}\lr{c^2+\frac{g_{\phi\phi}}{r^2}\frac{2K^{\rm KN}_{\rm rot}}{m}}}
}
{c^2+\frac{g_{\phi\phi}}{r^2}\frac{2K^{\rm KN}_{\rm rot}}{m}}\right]^2.\label{KKN5}
\end{split}
\end{equation}
In deriving the equation above 
we used the fact that, from \eqref{ds2KN}, 
\begin{equation}
c^2d\tau^2=-g_{tt}dt^2-2\frac{g_{t\phi}}{r}\sqrt{\langle v^2_{\phi}\rangle_{p}}dtd\tau-\frac{g_{\phi\phi}}{r^2}\langle v^2_{\phi}\rangle_{p}d\tau^2.
\end{equation}
It follows that
\begin{equation}
\lim_{r\rightarrow\infty}K_{\rm rot}^{\rm KN, t}=\frac{1}{2}\frac{mc^2}{\sin^2\theta+mc^2\beta}.
\end{equation}
Observe that on the plane $z=0$ ($\theta=\pi/2$) and in the classical regime $mc^2\beta >>1$ one obtains again $\lim_{r\rightarrow\infty}K_{\rm rot}^{\rm KN, t}=1/2\beta$, while a strongly relativistic system $mc^2\beta\sim1$ leads to $\lim_{r\rightarrow\infty}K_{\rm rot}^{\rm KN, t}\sim mc^2/4$.

\begin{figure}[h]
\hspace*{-0cm}\centering
\includegraphics[scale=0.26]{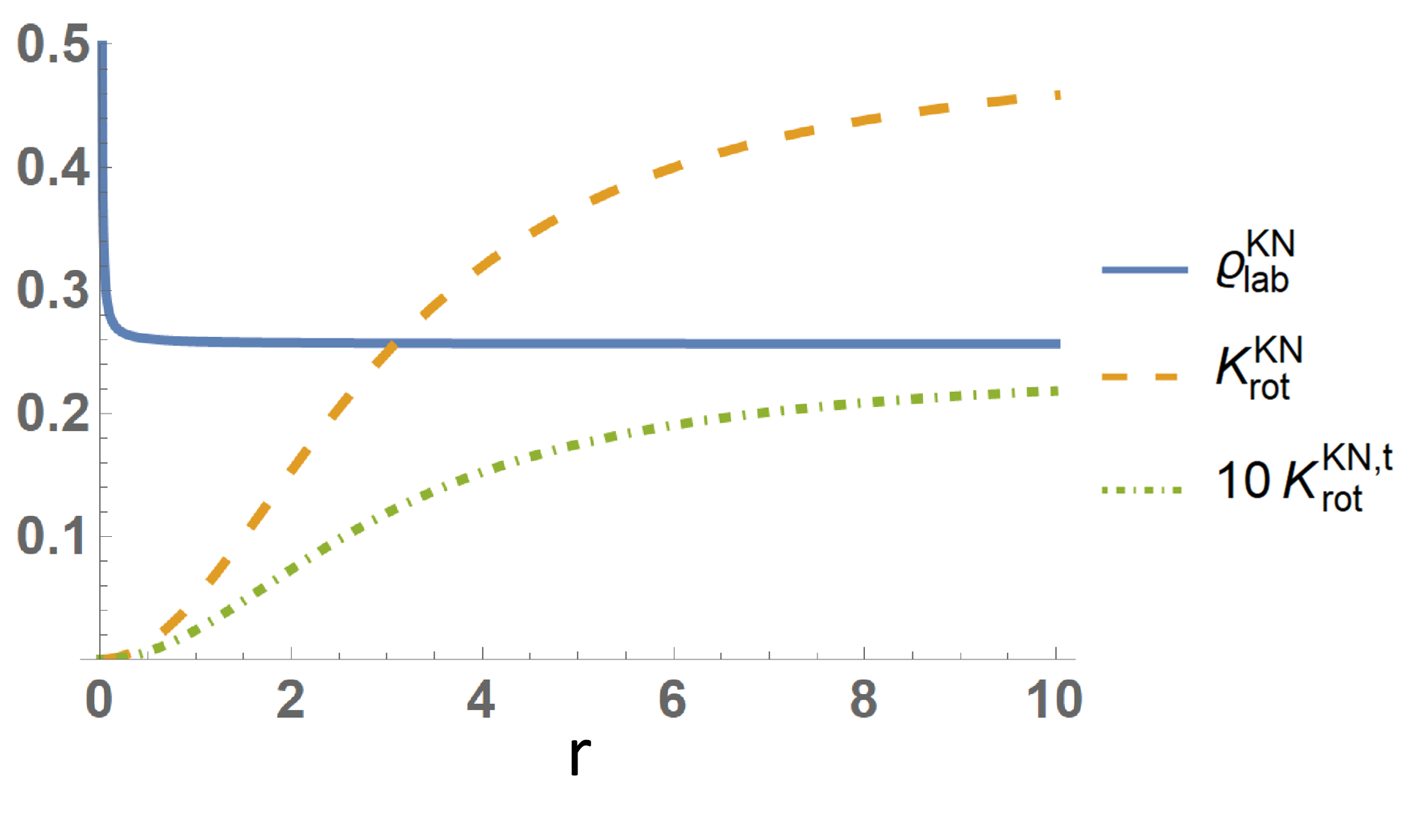}
\caption{\footnotesize Radial profiles of laboratory density \eqref{rhoKNlab}, 
rotational kinetic energy in proper time \eqref{KKN4}, and rotational kinetic energy in time $t$ \eqref{KKN5} at $z=0$ for $m=0.05$, $c=1$, $R_s=0.01$, $a=3$, $R_Q=0.01$, $p_t=-mc^2$, $\beta=1$, $e=1$, $G=1$, and $\epsilon_0=1$. Density is given in arbitrary units, while $K^{\rm KN,t}_{\rm rot}$ is scaled by a factor $10$.} 
\label{fig5}
\end{figure}

Figure \ref{fig5} shows radial profiles of Kerr-Newman laboratory density \eqref{rhoKNlab}, 
rotational kinetic energy in proper time \eqref{KKN4}, and
rotational kinetic energy in time $t$ \eqref{KKN5} on the plane $z=0$  
for the parameter values of the second example in figure \ref{fig4}.  
Observe that the rotational kinetic energies $K_{\rm rot}^{KN}$ and $K_{\rm rot}^{KN,t}$
are increasing functions of the radial coordinate, and they eventually converge toward 
a constant value, even though 
the particle density $\rho_{\rm lab}^{\rm KN}$ is a decreasing function of $r$.
 
\section{Concluding Remarks}

In this paper, we studied the effect of nontrivial spacetime metrics on 
statistical distributions. This problem arises, for example, 
when the effects of general relativity cannot be neglected in the description of single particle dynamics.

The formulation of statistical mechanics in the context of general relativity
represents a challenge because, in addition to the difficulty posed by the curvature of spacetime, 
the role of time, which affects temperature and thermodynamic equilibrium, is different from the classical one. 
The present theory relies on the assumption that the metric coefficients, and thus the geodesic Hamiltonian, 
are symmetric with respect to coordinate time $t$.
This hypothesis stems from the expectation that, if a system possesses an equilibrium state, 
all physical observables, including the spacetime metric, should eventually be independent of time $t$. 
This assumption implies that 
the geodesic equations of motion can be cast in the form of a 
6-dimensional canonical Hamiltonian system in proper time $\tau$ on the level set of the constant of motion $p_0$ arising from the
time-symmetry of the geodesic Hamiltonian. 
Then, the equilibrium distribution function is obtained by enforcing the ergodic hypothesis 
on the reduced phase space, and   
thermodynamic equilibrium is characterized by the property that  
the spatial particle distribution becomes a function of only three (spatial) coordinates. 
In general, spacetime curvature affects the particle equilibrium density distribution 
through the determinant of the spatial part of the metric tensor, which is related to
the Riemannian curvature tensor, and through an exponential factor 
where the spacetime components of the metric tensor appear. 

The construction above has been applied to Schwarzschild and Kerr-Newman spacetimes. 
In Schwarzschild spacetime, the effect of the metric tensor has been studied 
by taking into account the possibility that macroscopic constraints, 
such as angular momentum, may characterize the evolution of the ensemble.  
These constraints introduce nonlinearity in the relationship between the radial position of the peak in azimuthal rotation velocity, 
and the position $R_s$ of the event horizon of the source of the metric. 
Therefore, by appropriately tuning physical parameters, it is possible to achieve configurations in which
a decreasing density and a non-decreasing azimuthal rotation velocity coexist at radii much larger than 
the Schwarzschild radius $R_s$. 

In the Kerr-Newman configuration, the charge and rotation of the central mass
impart a heterogeneous structure to the particle distribution. 
In particular, we found that the rotational kinetic energy (the momentum-space average
of the squared modulus of azimuthal velocity)
becomes an increasing function of the radial coordinate, and eventually approaches
a constant value corresponding to classical equipartition of energy.
Furthermore, as in the Schwarzschild case, a decreasing spatial density 
does not imply a decreasing rotational kinetic energy.

\section*{Acknowledgment}
The research of NS was partially supported by JSPS KAKENHI Grant No. 17H01177.

\section*{Data Availability}

The data that support the findings of this study are available from
the corresponding author upon reasonable request.

\end{document}